\documentclass[pra,twocolumn,graphicx,showpacs,superscriptaddress]{revtex4}
\usepackage{epsfig,amsmath}

\begin{document}

\title{Electron-electron dynamics in laser-induced nonsequential double ionization}
\author{C. Figueira de Morisson Faria}
\email{faria@itp.uni-hannover.de}
\affiliation{Quantum Optics Group, Institut f\"ur Theoretische Physik, Appelstr. 2, Universit\"at Hannover, 30167 Hannover, Germany}
\affiliation{Max-Born-Institut, Max-Born-Str. 2A, D-12489 Berlin, Germany}
\author{X. Liu}
\email{xjliu@mbi-berlin.de}
\affiliation{Max-Born-Institut, Max-Born-Str. 2A, D-12489 Berlin, Germany}
\author{H. Schomerus}
\email{henning@mpipks-dresden.mpg.de}
\affiliation{Max-Planck-Institut f\"ur Physik komplexer Systeme, N\"othnitzer Str. 38, D-01187 Dresden, Germany}
\author{W. Becker}
\email{wbecker@mbi-berlin.de}
\affiliation{Max-Born-Institut, Max-Born-Str. 2A, D-12489 Berlin, Germany}
\date{\today}

\newcommand{\vecp}{\mathbf{p}}
\newcommand{\vecA}{\mathbf{A}}
\newcommand{\vecv}{\mathbf{v}}
\newcommand{\vecV}{\mathbf{V}}
\newcommand{\vecP}{\mathbf{P}}
\newcommand{\veck}{\mathbf{k}}
\newcommand{\vecx}{\mathbf{x}}
\newcommand{\vecr}{\mathbf{r}}
\newcommand{\half}{{\tiny \frac{1}{2}}}
\newcommand{\veca}{\mathbf{a}}

\begin{abstract}
For the description of nonsequential double ionization (NSDI) of rare-gas atoms by a strong linearly
polarized laser field, the quantum-mechanical $S$-matrix diagram that incorporates rescattering impact ionization is evaluated in the strong-field approximation. We employ a uniform approximation, 
which is a generalization of the standard saddle-point approximation.  We systematically analyze the manifestations  of the electron-electron
interaction  in the momentum distributions  of the ejected electrons: for the interaction, by which the returning electron frees the bound electron, we adopt either a (three-body) contact interaction or a Coulomb interaction, and we do or do not incorporate the mutual Coulomb repulsion of the two electrons in their final state. In particular, we investigate the correlation of the momentum components parallel to the laser-field polarization, with the transverse momentum components either restricted to certain finite ranges or entirely summed over.    In the latter case, agreement with experimental data is best for the contact interaction and without final-state repulsion. In the former, if the transverse momenta are restricted to small values, comparison of theory with the data shows evidence of Coulomb effects.  We propose that experimental data selecting events with small transverse momenta of \textit{both} electrons are  particularly promising in the elucidation of the dynamics of NSDI. Also, a classical approximation of the  quantum-mechanical $S$ matrix is formulated and shown to work very well inside the classically allowed region.
\end{abstract}
\maketitle

\section{Introduction}

Double and multiple ionization of atoms by intense laser fields is a very important process for laser-plasma diagnostics. 
As long as electrons are ripped off one by one, usually at the leading edge of the laser pulse while the intensity is rising \cite{augst}, the process can be straightforwardly described  in terms of rate equations and the Ammosov-Delone-Krainov (ADK) rates \cite{ADK}. However, as early as 1983 experimental evidence was found for the significance of a \textit{nonsequential} channel where several electrons are ejected in one coherent process \cite{lh83}. Nonsequential double and multiple ionization in an intense laser field is of great fundamental interest, since it requires electron-electron correlation as a necessary precondition. 
If one electron did not notice the other, all multiple ionization would be sequential. In most other intense-laser atom processes, such as high-order harmonic generation or above-threshold ionization, footprints of electron-electron correlation are hard to find \cite{footnmuller}. 
In contrast, in double ionization of helium below saturation, the nonsequential pathway was observed to be dominant by many orders of magnitude \cite{walker}.

The search for the physical mechanism that is capable of producing an effect of this magnitude remained inconclusive until experimental information became available that went beyond mere ion counting, that is, beyond \textit{total} double-ionization probabilities. The advent of the cold-target recoil-ion momentum spectroscopy (COLTRIMS) technique, also known as  reaction microscope \cite{coltrims}, combined with high-repetition-rate lasers, has opened the way towards acquisition of  \textit{multiply differential} cross sections  of the  double-ionization process. The first step was taken by the observation of the momentum distributions (all three components) of the doubly charged ion \cite{do00,mo00}. In principle, this technique enables one to record all six momentum components of two particles of opposite charge produced in some reaction. For double ionization, to the extent that the momentum transfer by the laser field is negligible, this amounts to complete kinematical characterization of the process. For a recent review, see Ref.~\cite{advan2}.

As a result, for the low-frequency high-intensity lasers that were employed in the COLTRIMS experiments, rescattering \cite{corkum} has emerged as a dominant mechanism.  This is the same mechanism that is responsible for high-order harmonic generation and high-order above-threshold ionization: an electron set free via tunneling is driven back by the field to its parent ion where it can rescatter, recombine, or dislodge another electron (or several electrons). Even though rescattering appears to be the dominant mechanism, many features of the data remain unexplained. An example is the behavior of the multiply differential cross sections near and below the classical threshold \cite{eremina}. This realm is inaccessible to classical methods, and the data are not compatible with the results of the quantum descriptions.

The unequivocal imprint on the data caused by the simplest version of rescattering -- that is, electron-impact ionization -- is a double-humped distribution 
of the momentum component $p_{\mathrm{ion}\parallel}$ of the 
doubly charged ion parallel to the (linearly polarized) laser field \cite{do00,mo00}. The humps are centered near $p_{\mathrm{ion}\parallel}\approx \pm 4\sqrt{U_P}$, where $U_P$ is the ponderomotive potential of the laser field \cite{pp}. This simple estimate  is a straightforward prediction of the rescattering model \cite{richard,classical}. 

Various routes have been followed in the theoretical description of nonsequential double ionization (NSDI). The most ab initio approach is the numerical solution of the time-dependent Schr\"odinger equation in three spatial dimensions for each electron \cite{taylor,muller}. Owing to the extreme demands on computing power, no explicit results have been obtained yet for the near-infrared frequencies used in experiments. A significant simplification  occurs if the inner electron is allowed to respond to the outer, but not vice versa \cite{crapola}.
Much more work has been done on the corresponding problem in one spatial dimension for each electron \cite{bauer,liu,dorr,lein,eberly}. 

An alternative quantum-mechanical approach has attempted to identify the dominant contribution to the $S$-matrix element for the NSDI process \cite{oldfaisal,kuchiev}, out of the multitude of diagrams that contribute. For low-frequency, high-intensity lasers, this has turned out to be the one that describes rescattering. It has been evaluated by several groups with different approximations \cite{abecker1,abecker2,popgor,richard,doublegoresl,popgorchaclo,nsdiuni,bede}. Of these, the evaluation with saddle-point methods \cite{popgor,doublegoresl,popgorchaclo,nsdiuni} requires the least computational effort and affords good physical insight.

Until recently, the theoretical effort concentrated on the computation of total NSDI rates. Only a few theoretical results exist for the differential yields that have been obtained with the help of the COLTRIMS method. These were obtained from the solution of the one-dimensional time-dependent Schr\"odinger equation \cite{lein}, by a classical analysis of excited two-electron configurations in a time-dependent electric field \cite{se01}, and by three-dimensional classical-trajectory methods \cite{beijing,beijing2}. 
 The latter have produced good agreement with those data that are sufficiently far in the classical regime. Other than that, thus far the calculation of  multiply differential yields has been the domain of the $S$-matrix approach. In the latter, a crucial element is the form of the electron-electron interaction by which the returning electron frees the second bound electron. This interaction is treated in the lowest-order Born approximation. The most natural choice appears to be  the Coulomb interaction. However, astonishingly, a three-body contact interaction at the position of the ion produces better agreement with the experimental data, at least for neon \cite{neonjpb}.

In this paper, we perform a systematic investigation of the rescattering contribution to the $S$-matrix element that describes NSDI. We calculate the distributions of the electron momentum components parallel to the laser field, and integrate  over the components perpendicular to the field either completely or partly. Experiments corresponding to the latter situation have been carried out recently. For example, the data were analyzed by binning the transverse momentum of the observed electron according to its magnitude \cite{resolvedar,resolvedne}. This is a further step towards the ultimate goal of kinematically complete experiments. 
 It should help one  disentangle mere phase-space effects from the nontrivial dynamics of electron-electron correlation and find clear 
signatures of the very interaction that is instrumental for the ejection of the second electron. We compare the two extreme limits of this crucial electron-electron interaction, namely the infinite-range Coulomb potential and the zero-range contact potential. In addition, we exactly implement the Coulomb repulsion between the two electrons in the final state \cite{repulsion} and study its effect on the afore-mentioned electron distributions.

We treat the problem in terms of quantum orbits \cite{orbits,science}. Such orbits are closely
related to the electron trajectories obtained classically within the
rescattering model. Their contributions are, however, superposed in the fashion of the quantum-mechanical path integral. Moreover, being complex they account for the electron tunneling from its initial bound state into the continuum owing to the action of the laser field. Indeed, the quantum-orbit
approach is capable of describing interference effects, and it remains 
applicable in energy regions that are out of reach to classical methods, where  the rescattering process is classically
forbidden (in a sense to be defined below in Sec.~\ref{saddlep}). Furthermore, this approach is computationally much less demanding and more transparent
than other quantum-mechanical treatments.

Our computations are performed within a specific uniform approximation \cite{atiuni,nsdiuni},
which is a generalization of the standard saddle-point approximation, widely
used in the context of atoms in strong laser fields \cite{science}.
The standard saddle-point
approximation requires that the saddles be well isolated,  whereas the uniform approximation in question only needs 
the saddles to occur in pairs, regardless of their separation. The latter condition is always satisfied by the
quantum orbits which occur in intense-laser-atom processes \cite{orbits}. 
In fact, for a specified final state, 
contributing orbits always come in pairs, one having a longer travel time than the other. At the
boundaries between  classically allowed and forbidden energy regions, these two orbits almost coalesce. Such a boundary causes a ``cutoff'' in the energy spectrum, such that the yield decreases steeply when the associated parameter proceeds into the classically forbidden region.

We also present and evaluate a very simple model to describe rescattering impact ionization that is classical in the following sense: The first electron enters the continuum by  quantum-mechanical tunneling, which is described by a rate formula that is highly nonlinear in the applied field, such that ionization predominantly takes place near the maxima of the field. The process of impact ionization is governed by the square of the form factor that occurs in the quantum calculation. Everything else is described in classical terms. Under the condition that the kinetic energy of the returning electron be much larger than the second ionization potential, the results of this classical model agree extremely well with the quantum-mechanical results.

The paper is organized as follows. In the next section, we provide the
necessary theoretical background, namely the transition amplitude for NSDI
in the strong-field approximation (Sec. \ref{transampl}) and the 
saddle-point and uniform approximations (Sec. \ref{saddlep}). This formalism is
subsequently applied to compute momentum distributions of electrons for the
contact or Coulomb interaction, including or not including the final-state electron-electron
repulsion (Secs. \ref{norep} and \ref{e-erep}, respectively). In Sec.~\ref{classicalmodels} we formulate and evaluate the just mentioned classical model of rescattering-induced NSDI.  In Sec. \ref{sec:exp} we relate our results to the existing experiments, and in the concluding Section \ref{concl} we  assess the merits and shortcomings of our approach.

\section{Background}

\subsection{Transition amplitude}

\label{transampl}In the strong-field
approximation (SFA), the transition amplitude for the NSDI process, caused by laser-assisted inelastic
rescattering, is given by \cite{oldfaisal,richard}
\begin{eqnarray}
M=-\int_{-\infty }^{\infty }dt\int_{-\infty }^{t}dt^{\prime }\langle
\psi _{\vecp_{1}\vecp_{2}}^{(\mathrm{V})}(t)| V_{12}U_{1}^{(\mathrm{V})}(t,t^{\prime })V_1\nonumber\\
\otimes U^{(0)}_2(t,t')|\psi _{0}(t^{\prime })\rangle , \label{rescatt}
\end{eqnarray}
where $V_1$ and $U_1^{(\mathrm{V})}(t,t^{\prime })$ denote the atomic
binding potential and the Volkov time-evolution operator acting on the first 
electron, $U^{(0)}_2(t,t')$ is the field-free propagator acting on the second electron, and $V_{12}$ is the electron-electron interaction through which the second electron is
freed by the first. Equation (\ref{rescatt}) can be interpreted as
follows. Initially, both electrons are bound in their 
ground state $| \psi _{0}(t^{\prime })\rangle$, which is approximated by the product 
$|\psi
_{0}^{(1)}(t^{\prime })\rangle \otimes | \psi
_{0}^{(2)}(t^{\prime })\rangle$ of the 
one-electron ground states $| \psi _{0}^{(n)}(t^{\prime
})\rangle =e^{i|E_{0n}|t^{\prime }}| \psi _{0}^{(n)}\rangle$ with ionization potentials $|E_{0n}|$. 
At the time $t^{\prime }$, the first electron is released through
tunneling ionization, whereas the second electron remains bound in its initial state.
Subsequently, the first electron propagates in the continuum, gaining energy from the field. At the later time $t$, it undergoes an 
inelastic collision with its parent ion, dislodging the second electron in this process$.$ The
final electron state is taken either as the product state of one-electron
Volkov states, or as a two-electron Volkov state \cite{repulsion}, with asymptotic momenta 
$\vecp_{1}$ and $\vecp_{2}$. The two-electron Volkov state exactly accounts for the electron-electron Coulomb repulsion, in addition to the interaction with the laser field. 

The SFA is commonly made in semi-analytical calculations of laser-atom processes effected by high-intensity low-frequency laser fields. Briefly, it consists in neglecting the influence of the binding
potential on the propagation of the electron in the continuum, and the action of the laser field on the bound electron. In addition, there are other physical ingredients of the exact transition amplitude that are \textit{not} part of the approximation (\ref{rescatt}): for example, when the first electron tunnels out or when it returns to the ion or in both instances, it may promote the bound electron into an excited bound state, from which the latter will tunnel out at a later time. 
This process tends to produce electron with low momenta, as discussed below. 
Moreover, except in the initial wave function, the presence of the ion is not accounted for.  

Expanding the Volkov time-evolution operator in terms of Volkov states,
\begin{equation}
U^{(\mathrm{V})}(t,t^{\prime })=\int d^3\veck |\psi^{(\mathrm{V})}_\veck(t)\rangle \langle \psi^{(\mathrm{V})}_\veck (t')|, \label{volkovexp}
\end{equation}
where
\begin{eqnarray}
\langle \vecr|\psi^\mathrm{(V)}_\vecp(t) \rangle = (2\pi)^{-3/2}\exp\{i[\vecp +\vecA(t)]\cdot\vecr\}\notag\\
\times \exp\left( -i\int^td\tau [\vecp +\vecA(\tau)]^2\right) ,\label{volkovstate}
\end{eqnarray}
the amplitude (\ref%
{rescatt}) can be written as 
\begin{equation}
M=-\int_{-\infty }^{\infty }dt\int_{-\infty }^{t}dt^{\prime }\int d^{3}\veck V_{%
\mathbf{pk}}V_{\mathbf{k}0}\exp [iS_\vecp(t,t^{\prime },\mathbf{k}%
)],  \label{prescatt}
\end{equation}
with the action 
\begin{eqnarray}
S_\vecp(t,t^{\prime },\mathbf{k})=-\frac{1}{2}\left[
\sum_{n=1}^{2}\int_{t}^{\infty }d\tau [\mathbf{p}_{n}+\mathbf{A}(\tau )]^{2}\right.\nonumber\\
\left.+\int_{t^{\prime }}^{t}d\tau [\mathbf{k}+\mathbf{A}(\tau )]^{2}\right]
+|E_{01}|t^{\prime }+|E_{02}|t .  \label{action}
\end{eqnarray}
Here $\mathbf{A}(t)$ denotes the vector potential of the laser field, $\mathbf{p}\equiv (\vecp_1,\vecp_2)$ the final electron momenta, and $\mathbf{k}$ the drift momentum of the first electron in between ionization and recollision. We use  the length gauge, and we  employ atomic units throughout. The binding
potential $V_1$ of the first electron and  the electron-electron interaction $V_{12}$ enter through their form factors 
\begin{equation}
V_{\mathbf{pk}}=\langle \mathbf{p}_{2}+\mathbf{A}(t),\mathbf{p}_{1}+
\mathbf{A}(t)| V_{12}| \mathbf{k}+\mathbf{A}(t),\psi
_{0}^{(2)}\rangle \label{vpk}
\end{equation}
and 
\begin{equation}
V_{\mathbf{k}0}=\langle \mathbf{k}+\mathbf{A}(t^{\prime })|
V_1| \psi _{0}^{(1)}\rangle .
\end{equation}
In this paper, for the binding potential $V_1$ we choose a Coulomb potential and for the wave functions $\psi_0^{(i)}(\vecr)$ ground-state hydrogenic wave functions.

\subsection{Saddle-point and uniform approximations}
\label{saddlep}
For sufficiently low  frequencies and high laser intensities,
Eq. (\ref{prescatt}) can be evaluated to a good approximation by the method of steepest
descent, which we will also refer to  as the saddle-point approximation. Thus, we must determine the values of $\mathbf{k}$, $t^{\prime }$ and $t$
for which $S_\vecp(t,t^{\prime },\mathbf{k})$ is stationary, so that  its partial derivatives with respect to these variables vanish. This
condition gives the equations
\begin{subequations}\label{seq}
\begin{eqnarray}
\left[ \mathbf{k}+\mathbf{A}(t^{\prime })\right] ^{2}&=&-2|E_{01}|,
\label{saddle1}\\
\sum_{n=1}^{2}\left[ \mathbf{p}_{n}+\mathbf{A}(t)\right] ^{2}&=&
\left[ \mathbf{k}+\mathbf{A}(t)\right] ^{2}-2|E_{02}|,  \label{saddle2}\\
\int_{t^{\prime }}^{t}d\tau \left[ \mathbf{k}+\mathbf{A}(\tau )\right] &=&0.
\label{saddle3}
\end{eqnarray}
\end{subequations}
Equations (\ref{saddle1}) and (\ref{saddle2}) express  energy conservation at the
ionization and rescattering times, respectively, while Eq. (\ref{saddle3}) determines
the intermediate electron momentum such that the first
electron returns to the ion. Obviously, the solutions $t'_s\ (s=1,2,\dots)$ of Eq.~(\ref{saddle1}) cannot be real. In consequence, then,  $t_s$ and $\mathbf{k}_s$ are complex, too. 

Equation (\ref{saddle2}) describes energy conservation in the rescattering process. From the point of view of the first electron, rescattering is inelastic, since it donates energy to the second electron. Let us ignore, for the moment, the ionization potential $|E_{01}|$ and consider linear polarization. Then, $\veck = -\vecA(t')$, and $\veck$ and $t'$ are real. For given $t'$, Eqs.~(\ref{saddle1}) and (\ref{saddle3}) then determine the rescattering time $t$ and the momentum $\veck$. In the space of the final momenta $\vecp=(\vecp_1,\vecp_2)$,  Eq. (\ref{saddle2}) is the equation of the surface of a six-dimensional sphere with its center at $(-\vecA(t),-\vecA(t))$ and its squared radius given by $[ \mathbf{k}+\mathbf{A}(t)] ^{2}-2|E_{02}|$. We only consider times $t'$ such that the latter is positive. Then all possible electron momenta that are classically accessible in the process where the first electron is ionized at the time $t'$ are located on the surface of this sphere. The union of all these spheres upon variation of $t'$ contains all final electron momenta that are in this sense classically accessible. Below, we will frequently refer to it as the ``classically accessible region''. Leaving this region along any path in the $(\vecp_1,\vecp_2)$ space, we experience a sharp ``cutoff'' in the yield. Quantum mechanics allows a nonzero yield outside the classically allowed region, which, however,  decreases exponentially with increasing distance from its boundary. Formally, this is accomplished by the fact that the exact solutions of the saddle-point equations (\ref{seq}), which are always complex, exhibit rapidly increasing imaginary parts. For a detailed analysis of the solutions of Eq.~(\ref{seq}) for the closely related case of above-threshold ionization, cf. Ref.~\cite{orbits}.

In the standard saddle-point method, the action (\ref{action}) in the matrix element (\ref{prescatt}) is expanded to second order about the solutions to the saddle-point equations (\ref{seq}), whereupon the integrations can be carried out with the result
\begin{subequations}
\label{sadresc}
\begin{eqnarray}
M^{(\mathrm{SPA})} &=&\sum_{s}A_{s}\exp (iS_{s}),  \label{sadresca} \\
S_{s} &=&S_{\mathbf{p}}(t_{s},t_{s}^{\prime },\mathbf{k}_{s}), \\
A_{s} &=&(2\pi i)^{5/2}\frac{V_{\mathbf{p}\mathbf{k}_{s}}V_{\mathbf{k}_{s}0}%
}{\sqrt{\det S_{\mathbf{p}}^{\prime \prime }(t,t^{\prime },\mathbf{k})|_{s}}}.
 \label{sadrescc}
\end{eqnarray}
\end{subequations}
Here the index $s$ runs over the \textit{relevant} saddle points, those that are visited by an appropriate deformation of the real integration contour, viz. the ($t,t',\veck$) plane, to  complex values, and  $S_{\mathbf{p}}^{\prime \prime }(t,t^{\prime},\veck)|_s$ denotes the
five-dimensional matrix of the second derivatives of the action (\ref{action}) with respect
to $t,t^{\prime }$ and $\mathbf{k}$, evaluated at the saddle-points. This approximation can only be applied
when all the saddle points are well isolated. However, as already mentioned, the saddle-point solutions come in pairs, whose two members approach each other very closely near the classical cutoffs, i.e.,  near the
boundaries of the classically allowed region. Furthermore, beyond the classical cutoffs, one
of the two saddle points would yield an exponentially increasing contribution. This saddle point is  not visited by the afore-mentioned deformed integration contour. Hence, this solution has to be discarded from the sum  (\ref{sadresca}). Such a procedure  leads
to cusps in the cutoff region, which are particularly problematic for
nonsequential double ionization. A detailed analysis of this problem is
given in Ref.~\cite{nsdiuni}.

In this paper, we will use a more general uniform approximation 
\cite{Bleistein}, which has been successfully applied  to 
above-threshold ionization \cite{atiuni}, high-order harmonic generation \cite{MB02}, and in an exemplary fashion to NSDI \cite{nsdiuni}.
 The uniform approximation is nearly as simple as the standard
saddle-point approximation, but much more powerful. It requires the same input as the former, namely the
amplitudes $A_{s}(s=i,j)$ and the actions $S_{s}(s=i,j)$ for each pair $(i,j)$
 of saddle-point solutions. 
In the classically allowed region, the
transition amplitude then is given by 
\begin{eqnarray}
M_{i+j} &=&\sqrt{2\pi \Delta S/3}\exp (i\bar{S}+i\pi /4)  \notag
\label{eq:unif1} \\
&&{}\times \left\{ \bar{A}[J_{1/3}(\Delta S)+J_{-1/3}(\Delta S)]\right. 
\notag \\
&&\left. {}+\Delta A[J_{2/3}(\Delta S)-J_{-2/3}(\Delta S)]\right\} ,  
\\
\Delta S &=&(S_{i}-S_{j})/2,\qquad \bar{S}=(S_{i}+S_{j})/2,  \notag \\
\Delta A &=&(A_{i}-iA_{j})/2,\quad \bar{A}=(iA_{i}-A_{j})/2. \notag
\end{eqnarray}
If the two saddle points are sufficiently far apart, the parameter $\Delta S$ is large, and the standard saddle-point approximation (\ref{sadresc}) can be retrieved with the help of the asymptotic 
 behavior 
\begin{equation}
J_{\pm \nu }(z)\sim \left( \frac{2}{\pi z}\right) ^{1/2}\cos (z\mp \nu \pi
/2-\pi /4)
\end{equation}
of the Bessel functions for large $z$ .

In the classically forbidden region, one of the saddles is avoided by  
the contour. This is accounted for by taking an appropriate functional
branch of the (multi-valued) Bessel functions, which will automatically be selected by
requiring a smooth functional behavior at a Stokes transition \cite{Berry,Henning1}. The transition occurs at 
\begin{equation}
\mathrm{Re}\,S_{\mathbf{p}}(t_{i},t_{i}^{\prime },\mathbf{k}_{i})=\mathrm{Re}
\,S_{\mathbf{p}}(t_{j},t_{j}^{\prime },\mathbf{k}_{j}).  \label{stokesline}
\end{equation}
and signifies that one of the saddles drops out of the steepest-descent integration contour. 
The energy position of such a transition approximately coincides with the
boundary between the classically allowed and forbidden energy regions.
Beyond the Stokes transition, 
\begin{eqnarray}
M_{i+j} &=&\sqrt{2i\Delta S/\pi }\exp (i\bar{S})  \notag \\
&&{}\times \left[ \bar{A}K_{1/3}(-i\Delta S)+i\Delta AK_{2/3}(-i\Delta S)%
\right] .\quad  \label{eq:unif2}
\end{eqnarray}
Again, the result of the saddle-point approximation may be recovered using the asymptotic
form  
\begin{equation}
K_{\nu }(z)\sim \left( \frac{\pi }{2z}\right) ^{1/2}\exp (-z)
\label{asympt2}
\end{equation}
for large $z.$ Inserting Eq. (\ref{asympt2}) into (\ref{eq:unif2}), it is
easy to show that only one saddle point contributes to the saddle-point
approximation in this energy region.

\section{Distribution of electron momenta}
\label{generalpic}
In the following, we will evaluate the matrix element $M_{i+j}$ [Eq.~(\ref{eq:unif1}) or (\ref{eq:unif2})]. We will restrict ourselves to the two shortest orbits, as explained below at the end of Sec.~\ref{secallow}. Its modulus square specifies the distribution of the asymptotic momenta of the two electrons generated in the process of NSDI. Following the analysis of the experimental data, we decompose the electron momenta into components parallel and perpendicular to the (linearly polarized) laser field, so that $\vecp_n \equiv (p_{n\parallel},\vecp_{n\perp})\ (n=1,2)$. In a typical reaction-microscope experiment, the momentum of one electron and the momentum of the doubly charged ion are measured (usually, it is not possible to record all six components). Thereafter, the momentum of the other (undetected) electron is calculated from the assumption of momentum conservation. Even in the case where all six components of the final electron momenta were known, plotting the results would require to integrate over some of them.  In most experiments, the momentum components (of the detected electron) transverse to the laser polarization are either not recorded at all, or binned into  certain intervals. Correspondingly, we will compute the momentum correlation function by either integrating entirely or partly over the transverse momenta. Hence, we shall calculate an integral of the type
\begin{equation}
D(p_{1\parallel},p_{2\parallel})=\int d^2 \vecp_{1\perp}d^{2}\vecp_{2\bot }|M_{i+j}|^{2},\label{eec}
\end{equation}
where the integration extends over some range of the final momenta, i.e. of their magnitudes and/or their relative orientation.

We consider the monochromatic linearly polarized laser field 
\begin{equation}
\vecA(t)=A_{0}\cos \omega t \,\hat\vecx,  \label{field}
\end{equation}
which satisfies $\vecA(t+T/2)=-\vecA(t)$ with $T=2\pi/\omega$. 
Electrons generated by a recollision event at a time within the interval $-T/4 \le t\le T/4$ (modulo $T$) tend to populate the third quadrant of the $(p_{1\parallel},p_{2\parallel})$ plane, while those from $T/4\le t \le 3T/4$ (modulo $T$) mostly populate the first quadrant. In each time interval, there are two contributing saddle-point solutions,  referred to above as  $i$ and $j$, which have to be added coherently in the matrix element 
$M_{i+j}$. If the laser intensity is sufficiently low, the two populations are practically disjoint. However, with increasing laser intensity, the classical boundaries expand, and the two populations begin to overlap significantly in the region where the  momenta $p_{1\parallel}$ and  $p_{1\parallel}$ are small.
 In this case, in principle, we have to superpose all 
four contributions coherently, viz., in Eq.~(\ref{eec})  we have to integrate $|M_{i+j}(-T/4 \le t \le T/4) + M_{i+j}(T/4 \le t \le 3T/4)|^2$.  Rather, we will neglect their interference by taking $|M_{i+j}(-T/4 \le t \le T/4)|^2 + |M_{i+j}(T/4 \le t \le 3T/4)|^2$. This simplifies the calculation significantly, because it  allows us to take advantage of the symmetry $|M(t,t',\vecp)|=|M(t+T/2,t'+T/2,-\vecp)|$. This procedure is justified 
by the fact that the relative phase between them is a rapidly oscillating
function. Indeed, we have checked for the case where $p_{1\parallel}=p_{2\parallel}$ that the exact and the approximate calculation produce virtually identical results, definitely so when the transverse momenta are integrated over.

\subsection{The ``classically allowed'' regime of parallel momenta}\label{secallow}

Rewritten in terms of the parallel and perpendicular momentum components $p_{n\parallel}$ and $ \vecp_{n\perp}$,  the saddle-point equations (\ref{seq}) read
\begin{subequations}\label{modseq}
\begin{eqnarray}
\left[ k+A_{0}\cos \omega t^{\prime }\right] ^{2}&=&-2|E_{01}|,\\
\sum_{n=1}^{2}\left[ p_{n\parallel }+A_{0}\cos \omega t\right] ^{2}&=&\left[
k+A_{0}\cos \omega t\right] ^{2}\notag\\
- 2|E_{02}|-\sum_{n=1}^{2}\vecp_{n\perp }^{2},
\label{s2}
\end{eqnarray}
with
\begin{equation}
\mathbf{k}=-\frac{1}{\omega (t-t^{\prime })}A_{0}(\sin \omega t-\sin \omega
t^{\prime })\hat\vecx \equiv k \hat\vecx.
\end{equation}
\end{subequations}
Equation~(\ref{s2}) defines a circle in the $(p_{1\parallel },p_{2\parallel })$
plane with its center  at $p_{1\parallel }=p_{2\parallel }=-A_{0}\cos \omega t$ and the square of its radius given by the right-hand side. For $\vecp_{1\perp}=\vecp_{2\perp}=0$, its interior is the projection onto the $(p_{1\parallel},p_{2\parallel})$ plane of the six-dimensional surface mentioned below Eqs.~(\ref{seq}). Inside any such circle,  the rescattering process is classically allowed. The radii decrease with increasing transverse kinetic energies of the final electrons. In effect, the transverse kinetic energies add to the second ionization potential $|E_{02}|$, up to the point where the classically allowed region shrinks to zero. 
Note that both the center and the radii of the circles defined above depend on the rescattering time $t$. The union of all circles defines  the entire classically allowed region in the $(p_{1\parallel},p_{2\parallel})$ plane \cite{hoeberly}. Depending on the intensity and the second ionization potential $|E_{02}|$, it may or may not include the origin $p_{1\parallel}=p_{2\parallel}=0$ \cite{nsdicutoffs}. 

The presence of the cutoffs at the boundary of the classical region and their dependence on $\vecp_{n\perp }$ pose a
serious problem for the application of the standard saddle-point approximation in
computations of momentum distributions. In fact, the integration over an
interval of transverse momenta will lead to a situation with many Stokes
transitions, whose energy positions vary continuously. As a direct
consequence, the artifacts coming from the breakdown of the saddle-point
approximation at the cutoffs will affect the resulting yield over a large
interval of longitudinal momenta $p_{n\parallel }$. Therefore, the uniform
approximation is not only a desirable, but a necessary tool for the computation of  the
momentum distributions for NSDI in terms of quantum orbits. This problem is
discussed in detail in Ref.~\cite{nsdiuni}.

For fixed final momenta, the saddle-point equations (\ref{modseq}) may have a large number of relevant solutions, which can be ordered by the length of their ``travel time'' Re($t-t'$).  
Below, we will consider the pair of the two shortest quantum orbits, i.e. those two having the shortest travel time. Due to spreading of the associated wave packets, usually these two make the dominant contributions \cite{newfoot}. Of these, the longer orbit is associated with a ``slow-down collision'', that is, an electron along this orbit is decelerated by the laser field when it is approaching the crucial collision with the bound electron. In classical one-dimensional model calculations, these orbits have been shown to be particularly efficient for NSDI  \cite{eberly}. A detailed discussion of these orbits is given in \cite{nsdiuni}. 

\subsection{No electron-electron repulsion in the final state}

\label{norep} In this subsection, we neglect the Coulomb repulsion of the two final-state electrons, so that the final state is
 the product state of one-electron Volkov states,  $| \psi
_{\vecp_{1}\vecp_{2}}^{(\mathrm{V})}(t)\rangle =| \psi
_{\vecp_{1}}^{(\mathrm{V})}(t)\rangle \otimes | \psi
_{\vecp_{2}}^{(\mathrm{V})}(t)\rangle $.

The form factor $V_{\mathbf{pk}}$ then is given explicitly by 
\begin{eqnarray}
V_{\mathbf{pk}}=\frac{1}{(2\pi )^{9/2}}\int  d^{3}\vecr_{1}d^{3}\vecr_{2}
e^{i[\mathbf{p}_{1}+\mathbf{A}(t)]\cdot \mathbf{r}_{1}}e^{-i[\veck+\vecA(t)]\cdot \vecr_1} \nonumber\\ \times e^{i[\mathbf{p}_{2}+
\mathbf{A}(t)]\cdot \mathbf{r }_{2}}V_{12}(\mathbf{r}_{1},\mathbf{r}_{2})\psi
_{0}^{(2)}(\mathbf{r}_{2}) + (\vecp_1 \leftrightarrow \vecp_2),  \label{formnorep}
\end{eqnarray}
where $V_{12}$ is the electron-electron interaction in question that is responsible for freeing the second electron.
 
Let us consider an  electron-electron interaction of the form
\begin{equation}
V_{12}\equiv V_{12}(\vecr_1,\vecr_2)= v_{12}(\vecr_1-\vecr_2)V_2(\vecr_2),\label{potdecomp}
\end{equation}
where $v_{12}(\vecr)$ only depends on the interparticle separation. 
Of course, in a truly microscopic description, there would be no potential $V_2$, but we may want to interpret $V_{12}$ as an \textit{effective} potential that incorporates the presence of the ion (which is positioned at the origin). 
Then,  the form factor (\ref{formnorep}) can be rewritten as
\begin{eqnarray}
V_{\mathbf{pk}}=\left[\tilde{v}_{12}(\vecp_1-\veck)+\tilde{v}_{12}(\vecp_2-\veck)\right]\nonumber\\
\times \int d^3\vecr_2 e^{-i[\vecp_1+\vecp_2-\veck-\vecA(t)]\cdot\vecr_2} 
V_2(\vecr_2)\psi^{(2)}(\vecr_2), \label{12symmetry}
\end{eqnarray}
where $\tilde{v}_{12}(\vecp)$ is the Fourier transform of $v_{12}(\vecr)$. Of course, $V_{\mathbf{pk}}$ is symmetric upon the exchange of $\vecp_1 \leftrightarrow \vecp_2$, but this does not hold if only individual components are interchanged, viz. $p_{1i}\leftrightarrow p_{2i}$ where $i=x,y$, or $z$. Only in the case where $v_{12}(\vecr)$ is of very short range, so that $\tilde{v}_{12}(\vecp)$ is constant, does this exchange symmetry hold component by component. On the other hand, this additional symmetry holds regardless of the shape of $V_2$. The effect of these symmetries will be encountered below.

 In the next two subsections, we will consider the two extreme cases
for the interaction $V_{12}(\vecr_1,\vecr_2)$,
the contact interaction with zero range and the Coulomb interaction with infinite range.

\subsubsection{Contact interaction}

\label{deltnorep} First, we investigate the three-body contact interaction 
\begin{equation}
V_{12}(\vecr_1,\vecr_2)=\delta (\mathbf{r}_{1}-\mathbf{r}_{2})\delta (\mathbf{r}_{2}),
\label{contact}
\end{equation}
which confines the electron-electron interaction to the position of the ion. 
For this interaction, the form factors $\ V_{\mathbf{pk}}$ and $V_{\mathbf{%
k}0}$ are constants. In this case and only in this case, one does not have to resort to the saddle-point approximation:  the  matrix element (\ref{rescatt}) can be obtained analytically up to
one quadrature \cite{richard,JMO03}. For any other potential, the exact evaluation  requires a numerical computation of  multiple integrals.

\begin{center}
\begin{figure}
\includegraphics[width=6cm,angle=0]{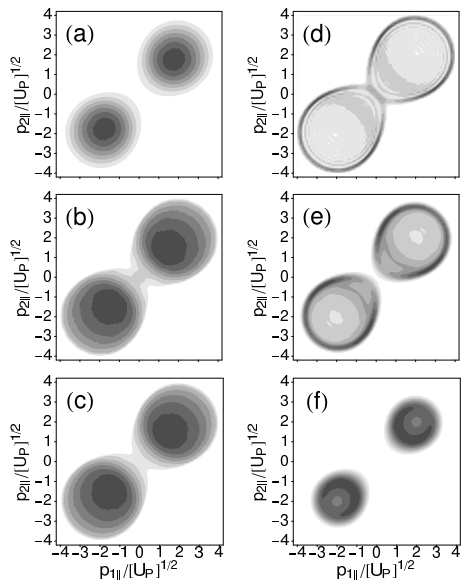}
\caption{Momentum correlation function (\ref{eec}) of the electron momenta parallel to the laser field  for nonsequential double
ionization computed with the uniform approximation using the contact
interaction (\ref{contact}). The field frequency is  $\protect
\omega =0.0551$ a.u. and the ponderomotive energy $U_{P}=1.2$ a.u., which corresponds to
an intensity of $5.5 \times 10^{14}\mathrm{W/cm}^{2}.$ The first two ionization
potentials  are  $
|E_{01}|=0.79$ a.u. and $|E_{02}|=1.51$ a.u. corresponding to
neon. Panel (a) shows the yield for the case where the transverse momenta $
\vecp_{n\perp}\ (n=1,2)$ are completely integrated over, whereas in the remaining panels they are  restricted to certain intervals. 
In panels (b) and (c), $\vecp_{2\bot }$ is integrated, while $0<p_{1\bot
}/[U_{p}]^{1/2}<0.1$ and $0.4<p_{1\bot }/[U_{p}]^{1/2}<0.5$, respectively.
In panels (d), (e), and (f), both transverse momenta are confined to the
intervals $0<p_{n\bot }/[U_{p}]^{1/2}<0.5$, $0.5<p_{n\bot }/[U_{p}]^{1/2}<1$,
and $1<p_{n\bot }/[U_{p}]^{1/2}<1.5,$ respectively.}\label{fig1}
\end{figure}
\end{center}

In Fig. \ref{fig1}, we display the momentum distributions (\ref{eec}) computed for this potential
with the uniform approximation, for various ranges of $|\vecp_{n\bot }|\ (n=1,2)$ and with the relative angle between $\vecp_{1\perp}$ and $\vecp_{2\perp}$ integrated over. 
In Fig.~\ref{fig1}(a), the transverse momenta are entirely summed over. The features
obtained, i.e., regions of circular shape around the two maxima at $\
p_{1\parallel }=p_{2\parallel }=\pm 2\sqrt{U_{p}}$, are in excellent
agreement with those in Ref. \cite{doublegoresl}.

The saddle-point equation (\ref{s2}) shows that the transverse kinetic energies add to the second ionization potential. In consequence, the higher the second ionization potential is and the lower the intensity, the more closely are the momentum correlation functions concentrated around the momenta $p_{1\parallel}=p_{2\parallel} = 2\sqrt{U_P}$. This effect can be verified by comparing panels (a) and (f): in (f) both transverse momenta are large such that the total transverse kinetic energy is between $U_P$ and $2.25 U_P$, while in (a), where all transverse momenta are summed over, the result is dominated by the contributions from the smaller ones. In panel (b), one transverse momentum is very small. Hence, the distribution is broader, as if the intensity were higher and/or the second ionization potential smaller than it actually is. 
Panels (d) and (e), where both transverse momenta are restricted to very small or moderately small values, have an appearance very different from the other panels.  The distributions are ring-shaped and concentrated near the boundary of the classically allowed region, while the region around $p_{1\parallel}=p_{2\parallel} = 2\sqrt{U_P}$ is almost unpopulated. 
Below, in Section \ref{classicalmodels}, we will be able to understand these features qualitatively as well as quantitatively from classical considerations.

\subsubsection{Coulomb interaction}

\label{coulnorep} In this subsection, we perform a similar analysis for the 
Coulomb interaction 
\begin{equation}
V_{12}(\vecr_1,\vecr_2)=\frac{1}{|\mathbf{r}_{2}-\mathbf{r}_{1}|}.  \label{Coulint}
\end{equation}
This appears to be a more realistic description of the
interaction, by which the first electron releases the second. In an ab-initio Born-series calculation, this interaction constitutes the lowest order. The ion is not accounted for, that is, the potential $V_2(\vecr_2)$ of Eq.~(\ref{potdecomp}) is absent. 
The corresponding form factor, 
\begin{eqnarray}
V_{\mathbf{pk}}\sim \frac{1}{(\mathbf{p}_{1}-\mathbf{k})^{2}[2|E_{02}|+(%
\mathbf{p}_{1}+\mathbf{p}_{2}-\mathbf{k}+\mathbf{A}(t))^{2}]^{2}}\nonumber\\
+\mathbf{p}%
_{1}\leftrightarrow \mathbf{p}_{2},  \label{formcoul}
\end{eqnarray}%
is a function of the electron velocities $\vecp_n+\vecA(t)$ and $\veck+\vecA(t)$ just after and just prior to, respectively, the crucial rescattering event. 

\begin{center}
\begin{figure}
\includegraphics[width=6cm,angle=0]{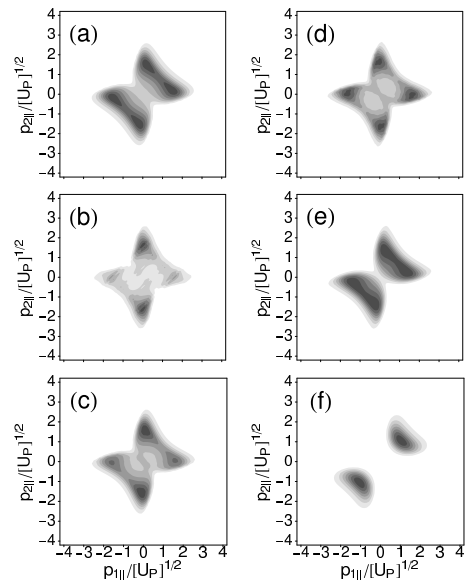}
\caption{Same as Fig.~\ref{fig1}, but calculated for the Coulomb interaction (\ref{Coulint}).}
\label{fig2}
\end{figure}
\end{center}

Using the uniform approximation, we again compute momentum distributions (\ref{eec}) for various transverse-momentum ranges regardless of the relative orientation of the transverse momenta. (We postpone showing a few distributions for \textit{fixed} relative angle till the very end of Sec.~\ref{generalpic}.)
These distributions are shown in Fig.~\ref{fig2}. The form 
factor (\ref{formcoul}) favors  small $\mathbf{p}_{1}-\mathbf{k}$ and/or small $\mathbf{p}_{2}+\mathbf{A}(t)$ [or small $\mathbf{p}_{2}-\mathbf{k}$ and/or small $\mathbf{p}_{1}+\mathbf{A}(t)$], which is equivalent with small momentum transfer of the returning electron to the bound electron and the bound electron being set free with small velocity \cite{footnote}. This means that the maxima of the distribution are
expected for small $\mathbf{p}_{1}$ and large $\mathbf{p}_{2} $ (or large $\mathbf{p}_{1}$ and small $\mathbf{p}_{2}$), i.e.,
away from the $\ p_{1\parallel }=p_{2\parallel }$ diagonal, since the vector
potential is near its maximum at the rescattering time $t$ while the drift momentum $\veck$ is small. When the transverse momenta are summed over  [Fig.~\ref{fig2}(a)], the result is in agreement with Ref.~\cite{doublegoresl}. In Figs. 2(b) and 2(c), we
restrict the transverse momentum of one of the electrons.  In comparison with the case of the contact interaction in Fig.~\ref{fig1}, this has a lesser effect on the momentum correlations. The signature of the Coulomb interaction -- one electron having a small and the other a large momentum -- is rather stable against summing over various parts of the transverse phase space. If both
transverse momenta are restricted [panels (d), (e), and (f)], we again 
observe, as in Fig.~\ref{fig1}, that the most significant contributions to the yield occur near the
boundaries of the classically allowed region, whose area decreases  for increasing
transverse momenta. The \textit{shape} of the distribution in panel (f), where both transverse momenta are large,  then does not look very different from Fig.~\ref{fig1}(f). It is, however,  concentrated at smaller values of $p_\parallel$. 

All panels of Figs.~\ref{fig1} and \ref{fig2} exhibit inversion symmetry with respect to the origin, that is, symmetry upon $(p_{1\parallel },p_{2\parallel })\leftrightarrow
(-p_{1\parallel },-p_{2\parallel })$. This is an immediate consequence of the symmetry $\vecA(t+T/2)=-\vecA(t)$ of a monochromatic laser field (\ref{field}). With the exception of Figs.~\ref{fig2}(b) and (c), all panels also show reflection symmetry  with respect to the diagonal 
$p_{1\parallel }=p_{2\parallel }$. Since the action (\ref{action}) is invariant upon interchanging all or some components of $\vecp_1$ and $\vecp_2$,  the presence or absence of this additional reflection symmetry is related to the corresponding symmetry properties of the form factor $V_{\vecp \veck}$, which are discussed below Eq.~(\ref{12symmetry}). In panels (b) and (c) of Figs.~\ref{fig1} and \ref{fig2}, the transverse momentum components of the \textit{detected electron} (electron 1) are binned, while those of the \textit{other electron} (electron 2) are summed over. For the contact interaction (\ref{contact}), $V_{\mathbf{pk}}$ is  constant and, therefore, trivially symmetric upon interchanging all or only some of the components of $\vecp_1$ and $\vecp_2$. This is not so for the Coulomb interaction (\ref{Coulint}). Hence, panels (b) and (c) of Fig.~\ref{fig1} do, and of Fig.~\ref{fig2} do not, exhibit reflection symmetry about the diagonal $p_{1\parallel }=p_{2\parallel }$. 

Panels (b) and (c) of Fig.~\ref{fig2} show that the longitudinal momentum of electron 1 (the one whose perpendicular momentum is restricted) has a higher propensity to be small than the same momentum component of electron 2, in violation of the reflection symmetry. This can be traced to the term $(\vecp_1-\veck)^{-2} = [\vecp_{1\perp}^2 +(p_{1\parallel}-k)^2]^{-1}$ of the form factor (\ref{formcoul}), the term that is related to the momentum transfer from the returning electron to the rest of the system. If $\vecp_{1\perp}^2$ is small, then the form factor is largest if $p_{1\parallel}$ is small as well, since the drift momentum  $k$ of the returning electron is small.

In principle, the presence or absence in the data of the reflection symmetry allows one to draw conclusions regarding the actual form of the interaction (\ref{potdecomp}). Indeed, experimental data that resolve the transverse momentum of the detected electron do show a violation of the $p_{1\parallel } \leftrightarrow p_{2\parallel }$ symmetry \cite{resolvedne,resolvedar}. However, there are experimental reasons that also lead to such a violation: the detector has a bias to detect the electron that arrives first, which is the faster one of the two electrons.

Another important conclusion derived from the comparison of Figs.~\ref{fig1} and \ref{fig2} is that the influence of the electron-electron interaction $V_{12}$ on the correlation functions is most pronounced if both transverse electron momenta are restricted to small values. Except for the fact that both respect the classical boundary, the distributions of Fig.~\ref{fig1}(d,e) on the one hand and Fig.~\ref{fig2}(d,e) on the other could hardly be more different. Notice, also, the dramatic difference between Fig.~\ref{fig1}(a) and Fig.~\ref{fig1}(d), while there is comparatively little difference between 
Fig.~\ref{fig2}(a) and Fig.~\ref{fig2}(d). 
For the contact interaction, which does not allow for any dynamics (apart from energy  conservation), phase space is the all-important feature, while for the Coulomb interaction the dynamical form factor overshadows the consequences of phase space. These facts combined  suggest that experiments for different rare gases with restricted transverse momenta \textit{of both electrons} might be best suited to unravel the differences between the electron-electron correlation in different atoms.

One should note that the results obtained for the Coulomb-type interaction are
strongly dependent on the gauge. Computations of NSDI yields in the velocity
gauge \cite{oldfaisal,abecker1,abecker2} give momentum distributions that are more concentrated near the diagonal $p_{1\parallel }=p_{2\parallel }$ and the
origin $p_{1\parallel }=p_{2\parallel }=0$. This is due to the fact that in the velocity gauge  the
form factor (\ref{formcoul}) is lacking the vector potential $\vecA(t)$ in the second factor of the denominator. Hence, the mechanism described above, which favors unequal momenta, is upset, and the form factor plainly  decreases for increasing momenta $\mathbf{p_{1}}$ and 
$\mathbf{p_{2}}$. The absence of the vector potential implies that the form factor does not depend on the instantaneous \textit{velocities} at the time of rescattering (as it does in the length gauge), but on the \textit{drift momenta}.

\begin{center}
\begin{figure}
\includegraphics[width=6cm,angle=0]{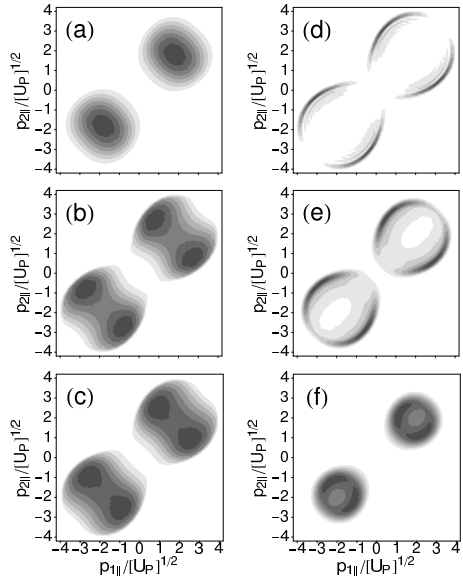}
\caption{Same as Fig.~\ref{fig1}, but including Coulomb repulsion of the two electrons in the final state. The transverse momenta of the two electrons are antiparallel, $\phi=\pi$.}
\label{fig3}
\end{figure}
\end{center}

\begin{center}
\begin{figure}
\includegraphics[width=6cm,angle=0]{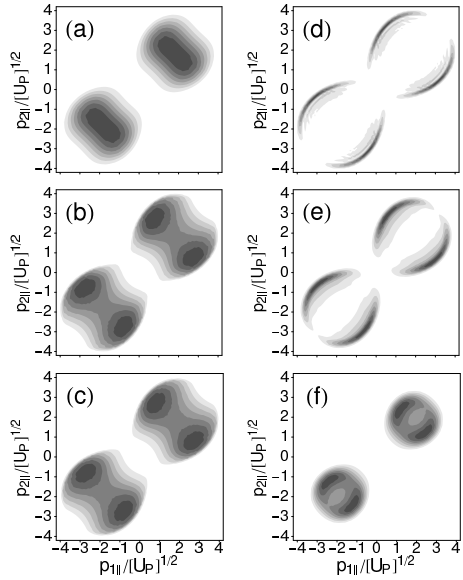}
\caption{Same as Fig.~\ref{fig3}, but the transverse momenta are at right angles, $\phi=\pi/2$.}
\label{fig4}
\end{figure}
\end{center}

\begin{center}
\begin{figure}
\includegraphics[width=6cm,angle=0]{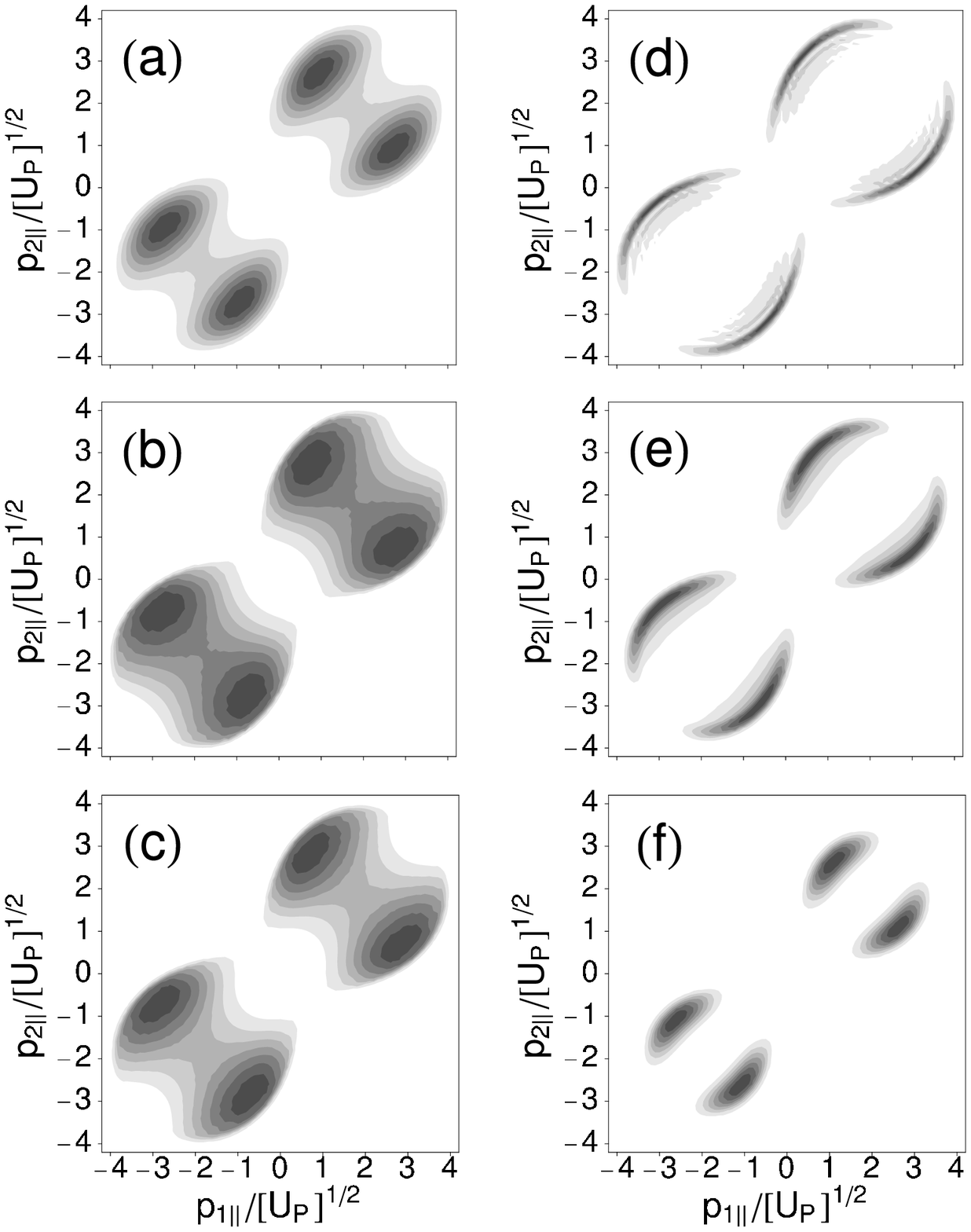}
\caption{Same as Fig.~\ref{fig3}, but the transverse momenta are parallel, $\phi=0$.}
\label{fig5}
\end{figure}
\end{center}

\begin{center}
\begin{figure}
\includegraphics[width=6cm,angle=0]{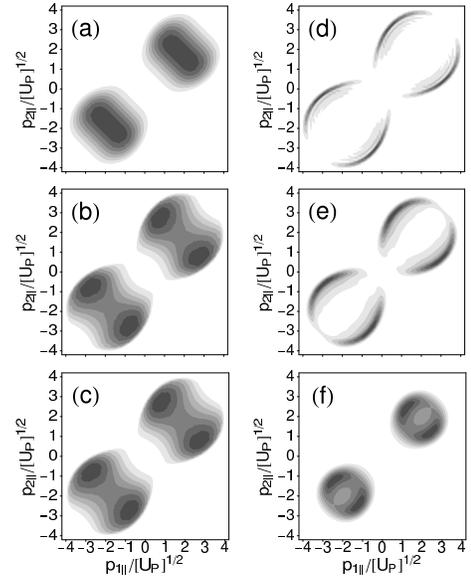}
\caption{Same as Fig.~\ref{fig3}, but the relative orientation of the transverse momenta is integrated over.}
\label{fig6new}
\end{figure}
\end{center}

\subsection{Electron-electron repulsion in the final state}

\label{e-erep} In this section, we take into account the repulsion of the two electrons in the final state. We do so by replacing in the matrix element (\ref{rescatt}) the product Volkov state by the exact correlated two-electron Volkov state whose wave function is \cite{repulsion}
\begin{eqnarray}
\Psi_{\vecp_1\vecp_2}^{(\mathrm{V,C})} (\vecr_{1},\vecr_{2},t)= \psi^{(\mathrm{V})}_{\vecp_1}(\vecr_1,t)\psi^{(\mathrm{V})}_{\vecp_2}(\vecr_2,t)\notag\\
\times\,  _{1}F_{1}(-i\zeta ;1;i(pr-\mathbf{p}\cdot\mathbf{r}))C(\zeta ),  \label{finalrep}
\end{eqnarray}
where $\vecr=\vecr_1-\vecr_2$, $\vecp=(\vecp_1-\vecp_2)/2$,
\begin{equation}
\zeta=|\vecp_1-\vecp_2|^{-1},
\end{equation}
and $_1F_1(a;b;z)$ denotes the confluent hypergeometric function. The normalization factor is
\begin{equation}
C(\zeta) = e^{-\pi\zeta/2}\Gamma(1+i\zeta),
\label{norm2}
\end{equation}
so that
\begin{equation}
|C(\zeta )|^{2}=\frac{2\pi \zeta }{\exp (2\pi\zeta)-1}.
\end{equation}%
The two-electron Volkov state has the simple form (\ref{finalrep}) since, owing to the dipole approximation,  the laser field couples only to the center of mass of the two electrons, while the Coulomb repulsion only affects their relative position. 
The prefactor (\ref{norm2}) will be found to have strong influence on the NSDI yields,
since it strongly favors unequal momenta.

The corresponding form factor $V_{\mathbf{pk}}$, originally defined in Eq.~(\ref{vpk}), is now to include the entire spatial part of the  two-electron Volkov function (\ref{finalrep}). Hence, in place of Eq.~(\ref{formnorep}) we have
\begin{eqnarray}
V_{\mathbf{pk}} =\frac{C^*(\zeta )}{(2\pi )^{9/2}}\int 
d^{3}\vecr_{1}d^{3}\vecr_{2}e^{i(\mathbf{p}_{1}-\mathbf{k})\cdot\mathbf{r}_{1}}
e^{i[\mathbf{p}_{2}+\mathbf{A}(t)]\cdot\mathbf{r}_{2}}\notag\\
\times V_{12}(\mathbf{r}%
_{1},\mathbf{r}_{2})\psi _{0}^{(2)}(\mathbf{r}_{2})   
\,  _{1}F_{1}(i\zeta ,1,-i(pr-\mathbf{p}\cdot\mathbf{r})).\label{formrep}
\end{eqnarray}

\subsubsection{Contact interaction}

For the contact interaction (\ref{contact}), the Coulomb-repulsion-modified form factor (\ref{formrep}) is just 
\begin{equation}
V_{\mathbf{pk}} \propto C^*(\zeta),\label{xx}
\end{equation}
which is directly proportional to the prefactor (\ref{norm2}). Note that it does not depend on the electron momentum in the intermediate state, but only on the final-state momenta, so that it can be taken out of all integrals in Eq.~(\ref{prescatt}). Its influence on the
momentum distributions is shown in the subsequent figures, in which the cases
of parallel, perpendicular, and antiparallel transverse momenta are 
investigated. 

Figure~\ref{fig3} deals with the case  of antiparallel transverse momenta, i.e., the electron momenta transverse to the field
polarization form an angle of $\phi=\pi$. For this angle, 
electron-electron repulsion is expected to play the least important role.
If the magnitudes of the transverse momenta are completely integrated over [Fig.~\ref{fig3}(a)], the
momentum distribution in the $(p_{1\parallel},p_{2\parallel})$ plane looks
very similar to the case without repulsion [cf., Fig.~\ref{fig1}(a)], except that it is slightly broader in the direction perpendicular to the diagonal  $p_{1\parallel }=p_{2\parallel }$. As one of the momenta is restricted to 
relatively low values [Fig.~\ref{fig3}(b)], each maximum near $\pm2\sqrt{U_p}$ splits
into two, which are positioned symmetrically with respect to 
$p_{1\parallel }=p_{2\parallel }$. As this momentum range is shifted to higher values,  
the two maxima start to merge
originating a plateau that extends across the diagonal [Fig.~\ref{fig3}(c)]. These features are physically expected, 
since Coulomb repulsion is more pronounced for small electron momenta.
The influence of  Coulomb repulsion can also be seen very clearly if both
transverse momenta are restricted to small values [Fig.~\ref{fig3}(d)]. Indeed, there
is a whole region around the diagonal  $p_{1\parallel }=p_{2\parallel }$ for
which the yield completely vanishes in comparison with the case without
repulsion [i.e., Fig.~\ref{fig1}(d)]. An analogous, less extreme effect is
present for slightly larger momenta [Fig.~\ref{fig3}(e)]. In fact, as compared to its
counterpart without repulsion [Fig.~\ref{fig1}(e)], there is a noticeable decrease in
the yield along and in the vicinity of $p_{1\parallel }=p_{2\parallel }$.
For large transverse momenta, on the other hand, Coulomb
repulsion makes hardly any difference [cf. Fig.~\ref{fig3}(f)].

The case  when the transverse momenta of the two electrons form a right angle, i.e., $\phi=\pi/2$ is intermediate (Fig.~\ref{fig4}). If the
transverse momenta are integrated over, the $(p_{1\parallel},p_{2\parallel})$-momentum distribution
considerably broadens in the direction perpendicular to 
$p_{1\parallel }=p_{2\parallel }$, as compared with the case without repulsion
and with the previous case. If one of the momenta is restricted to small values, the
distribution, again, exhibits the two distinct sets of maxima observed in the
antiparallel situation, with the main difference that such maxima are now more
pronounced and occur even if one of the momenta is not so small 
[e.g., in Fig.~\ref{fig4}(c)]. If both momenta are small, the yield looks almost identical with
that observed in the antiparallel case. As before, Coulomb repulsion has no noticeable effect, when both transverse momenta are large [panels (f)].

Finally, in Fig.~\ref{fig5} we address the most extreme situation, when both 
transverse momenta are parallel ($\phi=0$).
A general feature in this case is the sharp 
decrease in the yield near $p_{1\parallel }=p_{2\parallel }$, with two
distinct sets of maxima, symmetric with respect to 
$p_{1\parallel }=p_{2\parallel }$, for \textit{all} ranges of the transverse 
momenta, restricted or not.

Figure \ref{fig6new} shows the corresponding results when the relative angle $\phi$ is also integrated over. As expected, it looks much like an average of the momentum distributions of Figs.~\ref{fig3} -- \ref{fig5}.

\subsubsection{Coulomb interaction}

For the Coulomb interaction (\ref{Coulint}), the form factor $V_{\mathbf{%
pk}}$ can be evaluated with the help of the integral \cite{McDC}
\begin{eqnarray}
\int \frac{d^3\vecr}{r}e^{i\veca\cdot\vecr}\,_1\!F_1(i\nu;1;i(kr-\veck\cdot\vecr)) \notag\\
= 4\pi (\veca^2)^{i\nu-1}[(\veca-\veck)^2-\veck^2]^{-i\nu}.
\end{eqnarray}
This yields
\begin{eqnarray}
V_{\mathbf{pk}}\sim \frac{C^*(\zeta )}{(\mathbf{p}_{1}-\mathbf{k}
)^{2}[2|E_{02}|+(\mathbf{p}_{1}+\mathbf{p}_{2}-\mathbf{k}+\mathbf{A}
(t))^{2}]^{2}}\notag\\
\times \left[ 1-\frac{(\mathbf{p}_{1}-\mathbf{k})\cdot(\mathbf{p}_{1}-
\mathbf{p}_{2})}{(\mathbf{p}_{1}-\mathbf{k})^{2}}\right] ^{-i\zeta }+(\mathbf{p
}_{1}\leftrightarrow \mathbf{p}_{2}).  \label{formcoulrep}
\end{eqnarray}
Comparing this with the form factor (\ref{formcoul}) without final-state repulsion, we see that the former is now multiplied with the normalization factor (\ref{norm2}) as well as with the factor in square brackets. The latter now does depend on $\veck$, so it cannot, in principle, be pulled out of the integral. However, for given $\vecp$ in the classically allowed regime, the two saddle-point solutions $\veck_s$ of the respective pair of solutions are almost equal and, moreover, almost real. The contribution of the factor in square brackets in Eq.~(\ref{formcoulrep}) is then negligible as it is raised to a complex power. In the classically forbidden regime, its effect may be more significant, but in this regime the absolute yields are very small. All in all, the dominant effect of the final-state repulsion is due to the normalization factor $|C(\zeta)|^2$, as we observed already in Eq.~(\ref{xx}) for the contact interaction.
This has been confirmed by the identical results obtained taking both the exact
form factor (\ref{formcoulrep}) or the Coulomb form factor without repulsion multiplied by this factor (not shown).

\begin{center}
\begin{figure}
\includegraphics[width=6cm,angle=0]{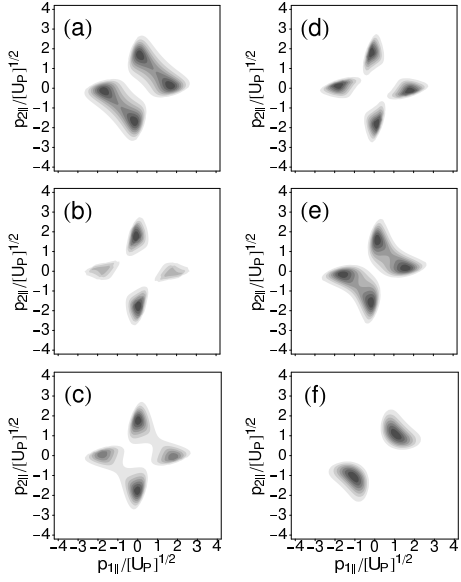}
\caption{Same as Fig.~\ref{fig1}, but calculated for the Coulomb interaction (\ref{Coulint}) and final-state Coulomb repulsion. The transverse momenta of the final electrons are antiparallel, $\phi=\pi$.}
\label{fig6}
\end{figure}
\end{center}

\begin{center}
\begin{figure}
\includegraphics[width=6cm,angle=0]{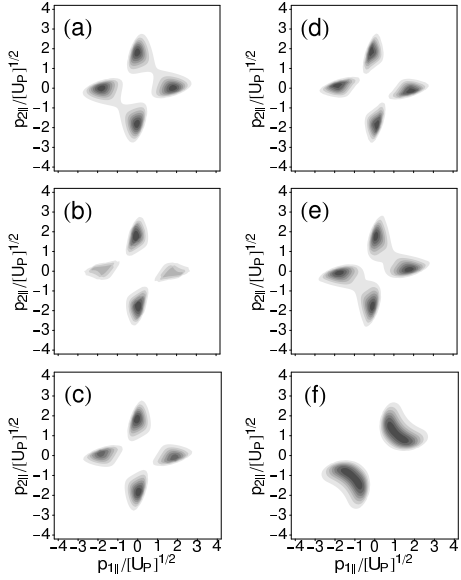}
\caption{Same as Fig.~\ref{fig6}, but the transverse momenta are perpendicular, $\phi=\pi/2$. }
\label{fig7}
\end{figure}
\end{center}

\begin{center}
\begin{figure}
\includegraphics[width=6cm,angle=0]{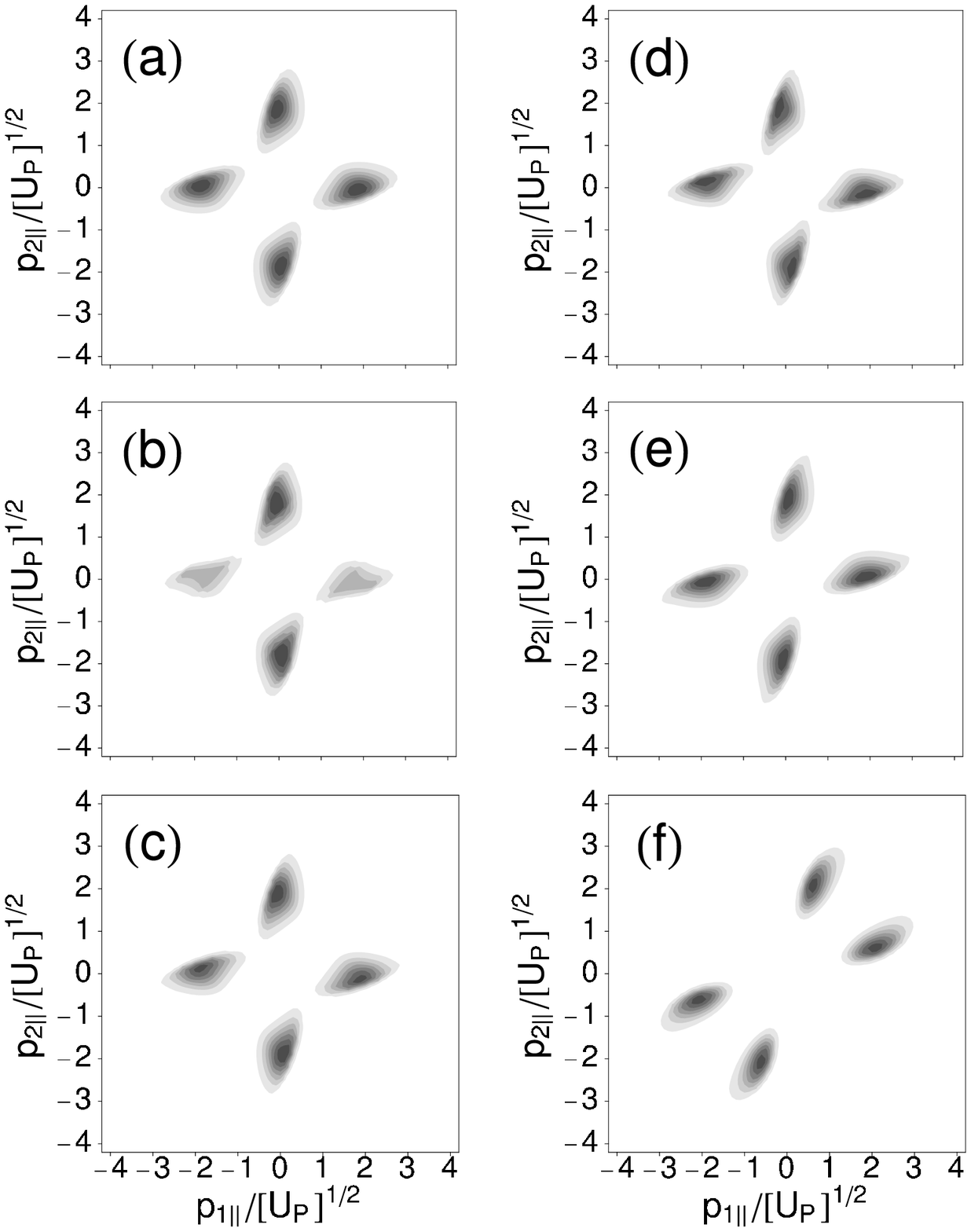}
\caption{Same as Fig.~\ref{fig6}, but the transverse momenta are parallel, $\phi =0$.}
\label{fig8}
\end{figure}
\end{center}

\begin{center}
\begin{figure}
\includegraphics[width=6cm,angle=0]{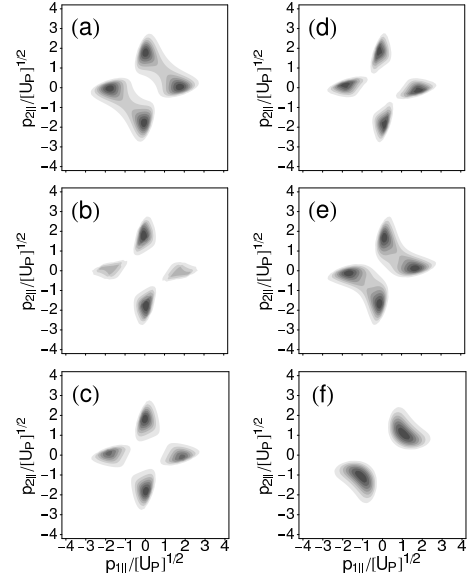}
\caption{Same as Figs.~\ref{fig6}--\ref{fig8}, but the relative orientation of the transverse momenta is integrated over.}
\label{fig10new}
\end{figure}
\end{center}
 
Once more, in Figs. \ref{fig6}--\ref{fig8}, we investigate NSDI for the transverse electron momenta being 
antiparallel, 
perpendicular,  and parallel, respectively.  The parallel case in Fig.~\ref{fig8} presents a very extreme 
example of the
influence of Coulomb repulsion: the momentum correlation function has shrunk to four spots, which are pushed away from the diagonal to the very edge of the classically allowed region. Figure \ref{fig8} should be compared with the corresponding results for the contact interaction in Fig.~\ref{fig5}, where this effect is much less dramatic, except in the case where \textit{both} transverse momenta are either large [panels (f)] or small [panels (d)]. 

The case where both transverse momenta are restricted to small values can be compared with two-electron one-dimensional model calculations. Indeed, momentum correlation functions calculated in this context from the numerical solution of the time-dependent Schr\"{o}dinger equation \cite{lein} look very much like those in panels (d) of Figs.~\ref{fig6}--\ref{fig8}. For these very small transverse momenta, their relative orientation hardly matters anymore, and the correlation function is concentrated in the four small regions $p_{1\parallel}=0,\, p_{2\parallel}=\pm 2\sqrt{U_P}$ and  $p_{2\parallel}=0,\, p_{1\parallel}=\pm 2\sqrt{U_P}$ on the axes. The very same feature can be observed in the results of Ref.~\cite{lein}.

Finally, in Fig.~\ref{fig10new} we present the results of integrating over the relative orientation $\phi$.

 Panels (b) and (c) of Figs. \ref{fig6} -- \ref{fig10new} again exhibit the lack of the $p_{1\parallel} \leftrightarrow p_{2\parallel}$ symmetry, which was discussed above in connection with Fig.~\ref{fig2}. As above the asymmetry is strongest in panels (b), where the transverse momentum $p_{1\perp}$ is restricted to the smallest values. The factor $|C(\zeta)|^2$, which incorporates final-state repulsion, is invariant upon $\vecp_1 \leftrightarrow \vecp_2$ component by component, since it depends on $|\vecp_1-\vecp_2|$. Therefore, the asymmetry is not affected when final-state repulsion is turned on.

\begin{center}
\begin{figure}
\includegraphics[width=8cm]{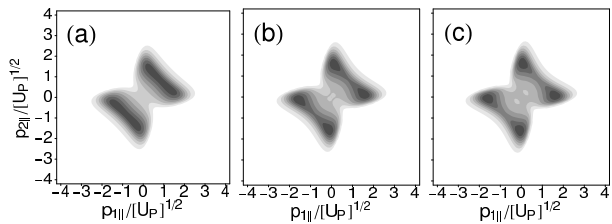}
\caption{Momentum correlation function for the Coulomb form factor (\ref{formcoul}) \textit{in the absence of final-state repulsion} for specific relative angles $\phi$: for $\phi=\pi$ (a), $\phi=\pi/2$ (b), and $\phi=0$ (c). The other parameters are as in Fig.~\ref{fig1}(a), in particular, the transverse momenta are completely summed over.}
\label{figaaa}
\end{figure}
\end{center} 

To conclude this Section, we investigate the momentum correlation as a function of the relative angle between the momenta of the final electrons \textit{in the absence of final-state repulsion}. For the contact interaction, is does not depend on this angle at all; for the Coulomb interaction (\ref{formcoul}), it is presented in Fig.~\ref{figaaa}. We only show the case, where the transverse momenta are entirely integrated. The dependence on the relative angle is weak: only a slight recess of population away from the diagonal is observed when the relative orientation turns from back-to-back to side-by-side [from panel (a) to (b)]. In the other cases (not shown), corresponding to panels (b) -- (f) of the previous figures, the dependence is similarly weak if not weaker.

\section{Classical models}\label{classicalmodels}
The saddle-point equations (\ref{seq}) pinpoint the crucial stages of NSDI: initial tunneling of the first electron, inelastic scattering, and free propagation in between these two events. Apart from the initial tunneling, the respective physics can largely be envisioned as classical, insomuch as the final electron momenta are classically accessible, and the better so the higher above threshold the inelastic rescattering takes place. Therefore, in this section we will explore a completely classical model.

Let us then consider the following expression for the NSDI yield (up to a constant factor) such that two electrons are generated with  drift momenta $\vecp_1$ and $\vecp_2$, 
\begin{eqnarray}
F(\vecp_1,\vecp_2) = \int dt' R(t') \delta \left(\half [\vecp_1+\vecA(t)]^2 \right.\nonumber\\
\left.+ \half [\vecp_2+\vecA(t)]^2 + |E_{02}| -E_{\rm ret}(t)\right) |V_{\vecp\veck}|^2\nonumber\\
=\int dt' R(t') \delta \left(\half (\vecp_{1\perp}^2+\vecp_{2\perp}^2) - \Delta E\right)|V_{\vecp\veck}|^2\label{app1}
\end{eqnarray}
with 
\begin{eqnarray}
\Delta E\equiv \Delta E(p_{1\parallel},p_{2\parallel},t) 
\equiv E_{\rm ret}(t)-|E_{02}|\nonumber\\
-\half [p_{1\parallel}+A(t)]^2 - \half [p_{2\parallel}+A(t)]^2. \label{deltaE}
\end{eqnarray}
Here the first electron appears in the continuum  with zero velocity at the time $t'$ according to the time-dependent rate $R(t')\equiv R(E(t'))$, for which we take $R(t')\sim |E(t')|^{-1}\exp \left[ -2(2|E_{01}|)^{3/2}/(3|E(t')|)\right]$ \cite{LL}. 
Starting from the position of the ion, the electron is accelerated by the laser field. The  time $t\equiv t(t') >t'$, at which the electron returns to the ion with kinetic energy $E_\mathrm{ret}(t) = (1/2)[\veck +\vecA(t)]^2$ is calculated classically along the lines of the simple-man model \cite{JPB94}. At this time, the electron  dislodges the second bound electron in an inelastic collision. The $\delta$-function in Eq. (\ref{app1}) expresses energy conservation in this inelastic collision. In fact, it is nothing but the saddle-point equation (\ref{saddle2}) with real $t,\,t'$ and $\veck$. The actual distribution of final momenta is governed by the form factor $|V_{\vecp\veck}|^2$, whose shape depends on the (effective) electron-electron interaction potential. 

Several features are absent in this model that are part of the quantum-mechanical description: (i)  spreading of the electronic wave packet from the ionization time $t'$ to the return time $t$. 
(ii) For given $\vecp_1$ and $\vecp_2$, there are several solutions $t\equiv t(t')$ (cf. the discussion at the end of Sec.~\ref{secallow}). In quantum mechanics, their contributions are added coherently in the amplitude, while in the total classical yield (\ref{app1}), 
the probabilities corresponding to the various solutions are added. 
(iii) Below the classical threshold, the argument of the $\delta$ function in Eq.~(\ref{app1}) is nonzero for any ionization time $t'$, and the yield is zero. Quantum mechanics admits larger energy transfer from the laser field to the charged particles, so that the yield remains nonzero, though it becomes exponentially small when the parameters move farther into the nonclassical regime. This implies that the classical model becomes already unreliable near the boundaries of the classical region.

We want to evaluate the distribution of the momentum components parallel to the laser field for particular configurations of the transverse components. This is governed by distribution functions of the type (\ref{eec}). 
In most cases, we are not interested in the relative orientation of the transverse momenta, and we restrict their magnitudes to certain ranges. This requires calculating
\begin{eqnarray}
D(p_{1\parallel},p_{2\parallel};P_1^2,P_2^2) = 2\pi \int_0^{P_1^2} dp_{1\perp}^2 \int_0^{P_2^2} dp_{2\perp}^2 \nonumber\\
\int_0^{2\pi} d\phi F(\vecp_1,\vecp_2). \label{distr1}
\end{eqnarray}
As in the quantum-mechanical considerations, we shall investigate the two extreme cases for the electron-electron potential $V_{12}$. 

\subsection{Contact interaction}

In this case, the form factor is a constant independent of the momentum of the returning electron as well as the momenta of the two final electrons. Therefore, the distribution of the final momenta is governed only by energy conservation at the instant of rescattering as expressed in the $\delta$ function in Eq.~(\ref{app1}), and by the available phase space. 
 The model is sufficiently simple that we can carry out integrations over the tranverse momenta analytically. To this end, we may, for example, replace the $\delta$ function by its Fourier transform,
\begin{equation*}
\delta (x) = \int_{-\infty}^\infty \frac{d\lambda}{2\pi} \exp (-i\lambda x).
\end{equation*}
Finite or infinite integrations over $\vecp_n$ can then be carried out straightforwardly, and the remaining integration over $\lambda$ is done with the help of \cite{GR}
\begin{equation}
\int^\infty_{-\infty}\frac{d\lambda}{(i\lambda+\epsilon)^{\nu}} e^{ip\lambda}=\frac{2\pi}{\Gamma (\nu)}p^{\nu -1}_+,
\end{equation}
where $x^\nu_+ = x^\nu \theta(\nu)$, with $\theta(\nu)$ the unit step function  and $\epsilon \rightarrow +0$.

This procedure yields
\begin{eqnarray}
D(p_{1\parallel},p_{2\parallel};P_1^2,P_2^2)=2\pi^2 \int dt' R(t') 
\left[2\Delta E \right. \nonumber\\ \left. +(2\Delta E-P_1^2-P_2^2)_+\right.\nonumber\\
\left.-(2\Delta E-P_1^2)_+ -(2\Delta E-P_2^2)_+\right].\label{distr2}
\end{eqnarray}
Note that this distribution is symmetric upon $p_{1\parallel}\leftrightarrow p_{2\parallel}$. Special cases include
\begin{equation}
D(p_{1\parallel},p_{2\parallel};\infty,\infty) 
 = 4\pi^2 \int dt' R(t')(\Delta E)_+,\label{distr3}
\end{equation}
where the transverse momenta are entirely integrated over [as in panels (a) of the figures of this paper],
\begin{equation}
D(p_{1\parallel},p_{2\parallel};P^2,\infty) = 4\pi^2 \int dt' R(t') \min\left[\half P^2,(\Delta E)_+\right],\label{distr4}
\end{equation}
where one electron is binned [cf. panels (b)], or
\begin{eqnarray}
D(p_{1\parallel},p_{2\parallel};P^2,P^2)=4\pi^2\int dt' R(t')\theta(\Delta E)\theta(P^2-\Delta E)\nonumber\\
\times \left[\Delta E \theta(P^2-2\Delta E) +(P^2-\Delta E)\theta(2\Delta E-P^2)\right].\label{distr5},
\end{eqnarray}
where both transverse momenta are restricted to the same range [cf. panels (d)]. The other distributions that we plotted can be obtained similarly.
 
\begin{center}
\begin{figure}
\includegraphics[width=6cm]{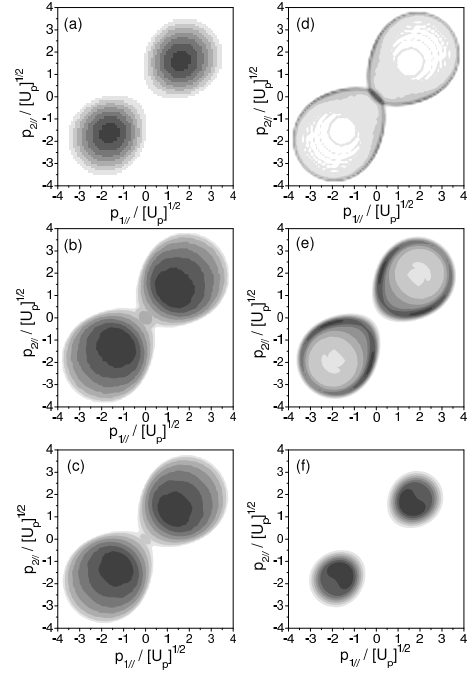}
\caption{Same as Fig.~\ref{fig1}, but calculated from the classical model for the contact interaction. Expression (\ref{distr3}) underlies panel (a), expression (\ref{distr4}) panels (b) and (c), and expression (\ref{distr5}) panel (d).}
\label{fig:clct}
\end{figure}
\end{center}

Momentum correlation functions calculated from Eqs.~(\ref{distr3}) -- (\ref{distr5}) are shown in Fig.~\ref{fig:clct}. Generally, they agree very well with the quantum-mechanical results of Fig.~\ref{fig1}. The minor differences that exist are most visible in the case where both transverse momenta are restricted to small intervals [panels (d) and (e)]. Here, the classical model emphasizes the boundary of the classical region projected onto the $(p_{1\parallel},p_{2\parallel})$ plane even more strongly than the quantum calculation. Of course, the latter extends into the classically forbidden region, but this is not visible on the scale of Fig.~\ref{fig1}.

The classical model and the expressions (\ref{distr2}) -- (\ref{distr5}) derived from it explain the dependence of the correlation-function distributions on the values of the transverse momenta and, in particular, the peculiar behavior visible in panels (d) and (e) of Figs.~\ref{fig1} and \ref{fig:clct}.  In order to satisfy the $\delta$-function condition in Eq.~(\ref{app1}), $\Delta E$ must be small (large) for small (large) transverse momenta. Let us consider the case of large momenta first. For given $t$, the quantity $\Delta E$ is largest around $p_{1\parallel}=p_{2\parallel}=-A(t)$, and, as a function of $t$, its absolute maximum is $\Delta E_\mathrm{max} \equiv 3.17U_P-|E_{02}|$.  Large transverse momenta are only possible near rescattering times corresponding to this maximum and, therefore, are concentrated around $p_{1\parallel}=p_{2\parallel}=\pm 2\sqrt{U_P}$ \cite{footn}. This is very visible in Fig.~\ref{fig:clct}(f). 
The integrated correlation function (\ref{distr3}) has its maximum at about the same momenta. However, it is broader, since it receives additional contributions from times $t'$, where $\Delta E(t)$ is smaller, as well as from smaller transverse momenta. 
 If one transverse momentum is binned with small values, the applicable distribution is given in Eq.~(\ref{distr4}). Comparing this with Eq.~(\ref{distr3}), we see that large 
$\Delta E$ are now less favored and, in consequence, the maximum of the distribution moves to lower values of $p_{n\parallel}$, and the distribution is still broader. This is clearly visible in panel (b) of Figs.~\ref{fig1} and \ref{fig:clct}.
When both transverse momenta are small such that $0\le |\vecp_{n\perp}|\le P$, the pertinent distribution (\ref{distr5}) shows that this requires $\Delta E \le P^2$. For Fig.~\ref{fig:clct}, $P^2=0.25$, so $\Delta E$ must be small. For times $t$ such that $E_{\mathrm{ret}}$ is much larger than $|E_{02}|$ this requires that  $p_{1\parallel}+A(t)$ and  $p_{2\parallel}+A(t)$ be large, which produces the ring-shaped population. There are contributions, too, from times $t$ where $E_{\mathrm{ret}}$ is not much larger than $|E_{02}|$. They populate the interior of the rings, but they are much weaker, since their ionization times are significantly below the maxima of $R(t')$. 
It is important to recall that the features just discussed  only depend on phase space and on the highly nonlinear form of the injection rate $R(t')$. 
Any deviation from the patterns depicted in Figs.~\ref{fig1} and \ref{fig:clct} is due to the form factor $V_{\vecp\veck}$ favoring certain momenta over others. Hence a comparison of the measured momentum correlations with those of Figs.~\ref{fig1} and \ref{fig:clct} does yield information about the actual electron-electron correlation mechanism.

With the method described above, arbitrary components of the final momenta can be summed over. In particular, single-electron momentum distributions in coincidence with NSDI can be computed by integrating over the momentum of one electron. This will not be pursued in this paper.

\subsection{Electron-electron Coulomb interaction}
The distribution function (\ref{distr1}) can also be evaluated analytically  in the presence of the Coulomb form factor (\ref{formcoul}), which depends on $p_{1\perp}^2$, $p_{2\perp}^2$, and the relative angle $\phi$ between $\vecp_1$ and $\vecp_2$.
For arbitrary upper limits $P_1^2$ and $P_2^2$, again all integrals (up to the integration over $t'$) can be carried out analytically: First, the integral over $\phi$ leads to a compact expression. Subsequently, the integration over $p_{2\perp}^2$  can be carried out trivially by means of the $\delta$ function in Eq.~(\ref{app1}), so that  
\begin{equation}
\frac12 (p_{1\perp}^2 + p_{2\perp}^2) -\Delta E =0, \label{delta0}
\end{equation}
with $\Delta E$ defined in Eq.~(\ref{deltaE}). The remaining integral over $p_{1\perp}^2$ then leads to a result that is too lengthy to be written down, but still analytical. The only integration that requires a numerical effort is the integration over the ionization time $t'$. Results of this procedure will  be presented elsewhere.

\subsection{Coulomb repulsion between the final-state electrons}
Finally, the Coulomb repulsion between the two electrons in the continuum can be incorporated by replacing
\begin{equation}
|V_{\vecp\veck}|^2 \rightarrow  |V_{\vecp\veck}|^2 |C(\zeta)|^2
\end{equation}
with the function $C(\zeta)$ from Eq.~(\ref{norm2}). This was exact for the contact potential [cf. Eq.~(\ref{xx})] and approximate for the Coulomb potential [cf. Eq.~(\ref{formcoulrep})]. Including this factor precludes, in general, performing the integration (\ref{distr2}), for the contact as well as for the Coulomb interaction, over the tranverse momenta in analytical form, owing to the functional form of $|C(\zeta)|^2$. There is one exception: if the final transverse momenta of the two electrons are perpendicular, $\vecp_{1\perp}\cdot\vecp_{2\perp}=0$, then $\zeta^{-2}=(p_{1\parallel}-p_{2\parallel})^2 + p_{1\perp}^2+p_{2\perp}^2$, and we have $p_{1\perp}^2+p_{2\perp}^2=\Delta E$  by Eq.~(\ref{delta0}). In this case, the function $|C(\zeta)|^2$, which cannot be  integrated in analytical form, actually does not have to be integrated over.

\begin{center}
\begin{figure}
\includegraphics[width=6cm]{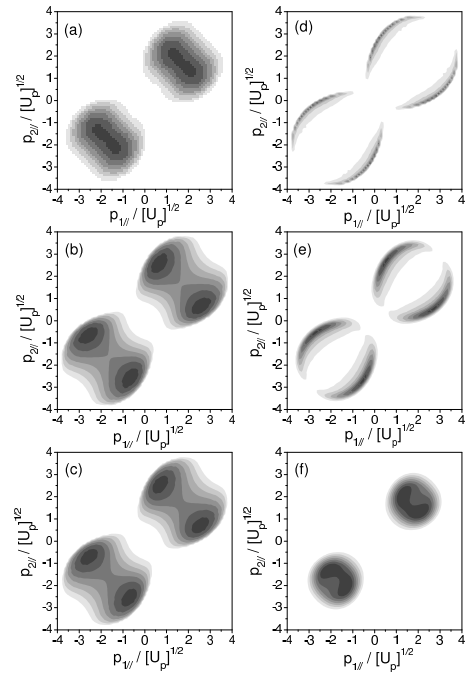}
\caption{Same as Fig.~\ref{fig:clct}, but including electron-electron repulsion in the final state. The transverse momenta are perpendicular, $\phi=\pi/2$.}
\label{fig:clres}
\end{figure}
\end{center}

The results of such a calculation for the case $\phi=\pi/2$ are presented in Fig.~\ref{fig:clres}. Comparison with the corresponding quantum-mechanical calculation in Fig.~\ref{fig7} again shows agreement well into small details. Virtually the only discrepancies are located near the diagonal, which the classical model clears of any population even more efficiently than the quantum-mechanical version. 

\section{Comparison with experimental data}\label{sec:exp}

The cold-target recoil-ion momentum spectroscopy (COLTRIMS) or reaction-microscope technique \cite{coltrims} allows, in principle, recording all three components of the momentum vectors of two particles with opposite charge ejected in some reaction process. In so much as the momentum imparted by the laser field can be neglected, this permits a complete kinematical analysis of laser-induced double ionization. Experiments so far have concentrated on the rare gases helium, neon, and argon. In a first round of experiments, the momentum of the doubly charged ion was registered for helium \cite{do00}, neon \cite{mo00}, and argon \cite{we00}. The second stage focused on the correlation of the two electrons \cite{nature,expe2,neonjpb,resolvedne,resolvedar,eremina}. Typically, the momentum components parallel to the (linearly polarized) laser field of one of the two electrons and of the ion  were recorded, regardless of the components perpendicular to the laser field. The momentum of the second electron is then inferred from momentum conservation. In Figs. \ref{fig1}--\ref{fig10new}, our results corresponding to such an analysis of the data are presented in the panels labeled (a).
The most detailed results are available for argon, for which  the correlation of the parallel components is analyzed  \cite{resolvedne,resolvedar}, while the transverse component of the momentum of the detected electron is binned into certain intervals. The theoretical results for such a  situation, but for the parameters of neon,  are displayed in panels (b) and (c) of Figs. \ref{fig1}--\ref{fig10new}. In the panels (d), (e), and (f), the transverse momenta of both electrons are confined to certain ranges. Such data have not been published yet. In a recent experiment \cite{bede}, the correlation of the transverse momenta was investigated, with the longitudinal components summed over.

In the experiments, characteristic differences have been established between NSDI of argon and helium on the one hand, and neon on the other. In argon, a significant number of events is found where the momenta $p_{1\parallel}$ and $p_{2\parallel}$ are either both small or such that they correspond to back-to-back emission, so that they come to lie in the second or fourth quadrant of the $(p_{1\parallel},p_{2\parallel})$ plane \cite{nature,resolvedne}. Very few such events are seen in neon \cite{neonjpb}. There is some consensus that these events are caused by the recolliding first electron exciting the second bound electron into an excited bound state from which it tunnels out at a later time \cite{expe2,finalfoot}. This mechanism is not part of the model considered in this paper (in one spatial dimension, it has been incorporated in Ref.~\cite{richard}). 
The different behavior of helium/argon versus neon has been attributed to the different energy dependence of the pertinent electron-ion cross sections for excitation and ionization of the respective ions \cite{jesus}.

For a detailed comparison between the  results of the models presented in this paper and the data, for precisely the conditions of the latter, we refer to Ref.~\cite{tobepublished}. In what follows, we will just compare the tendencies of our current results with those derived from the data.
For neon, the momentum correlation functions calculated for the contact potential and  integrated over all transverse momenta [Fig.~\ref{fig1}(a); see also Ref.~\cite{doublegoresl}] agree quite well with the data of Ref.~\cite{neonjpb} (and also with those of Ref.~\cite{resolvedne}; see below). Note that these theoretical results do not include the Coulomb repulsion in the final state. For the case where the transverse momentum of one electron is binned, data exist for argon only, while all of our calculations are for neon. However, our model does not crucially depend on the atomic species, and we plotted all momentum distributions on the scale of $p/\sqrt{U_P}$. Keeping in mind that the distributions broaden when the second ionization potential decreases \cite{richard}, 
we expect the tendencies that emerge in our results for neon to apply to argon as well. Inspecting, then, the argon data \cite{resolvedne} where one transverse momentum is binned  to the interval $0\le |\vecp_\perp| \le 0.5$ a.u., we notice a slight but distinct broadening of the distribution away from the diagonal $p_{1\parallel}=p_{2\parallel}$. This is similar to the tendency visible in panels (b) and (c) of Fig.~\ref{fig6new}, which do include the final-state Coulomb repulsion. Note that these data show no similarity with panels (b) and (c) of Figs.~\ref{fig10new}, which also include the final-state Coulomb repulsion, but are calculated for the case where $V_{12}$ is given by the Coulomb potential (\ref{Coulint}). However, the data are also compatible with panels (b) and (c) of Fig.~\ref{fig2}, which correspond to the Coulomb potential for $V_{12}$ and \textit{no} final-state repulsion.

Another set of electron-electron correlation data in argon \cite{resolvedar} has accomplished even tighter binning of the transverse momenta. Here, too, for $|\vecp_\perp|\le 0.3$ a.u. the tendency of the distributions to broaden away from the diagonal $p_{1\parallel}=p_{2\parallel}$ is obvious. For the very smallest bin, $0\le |\vecp_\perp|\le 0.1$ a.u. (Fig. 1a of Ref.~\cite{resolvedar}), the measured distribution in the  $(p_{1\parallel},p_{2\parallel})$ plane now does show a pattern with four well-separated maxima located on the $p_{1\parallel}$ and $p_{2\parallel}$ axes. This is  reminiscent of the panels (b) of Figs.~\ref{fig6}--\ref{fig10new}, which are calculated for about the same binning and intensity (though for neon) and include both the final-state repulsion and the Coulomb potential $V_{12}$. The contrast of the measured distribution, however, is much less pronounced than in Figs.~\ref{fig6}--\ref{fig10new}. All in all, the data agree better with a symmetrized version of Fig.~\ref{fig2}(b), which takes the Coulomb interaction for $V_{12}$, but does not include the final-state Coulomb repulsion.

It is remarkable that, apart from the case last mentioned, the data show little similarity with the model calculations that take the Coulomb repulsion for $V_{12}$. In no case do they agree with what one might expect to be the optimal description: the Coulomb potential for $V_{12}$ plus final-state Coulomb repulsion.

\section{Conclusions}

\label{concl} In this paper, we have performed a systematic investigation of the electron-electron dynamics  in non-sequential double
ionization within the strong-field-approximation framework. We have evaluated  the SFA
transition amplitudes by means of the uniform approximation \cite%
{nsdiuni,atiuni}, which, apart from being valid in all energy regions,
considerably simplifies the computations compared with a numerical
evaluation  \cite{abecker1,abecker2,bede} or the  solution of the time-dependent 
Schr\"{o}dinger equation \cite{taylor,muller}.

Our main concern is the effect of the electron-electron interaction on the
correlation of the electron-momentum components parallel to the polarization of the laser field, for the case where the transverse components are either  not detected at all or restricted to certain intervals. First, we ask the question of whether the
effective interaction $V_{12}$, which frees the second electron and is treated in lowest-order Born approximation, is of short range 
or long range. Second, we do include or we do not include the 
electron-electron Coulomb repulsion in the final two-electron state.

The results of such investigations are
at first sight very surprising: When the transverse momenta are integrated over, the apparently crudest approximation -- where the electron-electron interaction by which the second electron is  kicked out is treated as an effective three-body contact interaction,  and electron-electron repulsion in the final state is ignored -- yields the best agreement with the data. 
Comparison of the available data with the model calculations reveals some evidence of Coulomb effects only when one of the transverse momentum components is very small. Unfortunately, the available experimental data have not been analyzed to extract momentum correlations where \textit{both} transverse momenta are small. In this case, the four variants of the strong-field $S$-matrix model that we investigate -- three-body contact interaction vs. Coulomb interaction and Coulomb repulsion vs. no Coulomb repulsion in the final state -- exhibit the most pronounced differences. Owing to this high sensitivity, one is led to surmise  that such an analysis of the data might most clearly unveil the fundamental dynamics.

Another property of the correlation of the electron momenta $p_{n\parallel}$ parallel to the laser field that can be traced back to  the crucial electron-electron interaction $V_{12}$ is a lack of symmetry upon the interchange $p_{1\parallel}\leftrightarrow p_{2\parallel}$, as discussed below Eq.~(\ref{12symmetry}). It occurs when the two electrons are not treated on equal footing in the data analysis: the transverse momentum of the detected electron is binned while the other one is integrated over. The symmetry then is violated for the case where $V_{12}$ is given by a Coulomb potential and observed when it is contact interaction, regardless of whether or not the Coulomb repulsion between the final electrons is taken into account. However, there are also experimental causes unrelated to this fundamental reason that lead to a violation of the symmetry.

The most relevant aspect of the contact-interaction model, be it the quantum-mechanical $S$-matrix formulation or the classical version, might be its bare-bones character: arguably, there is no simpler model that accounts for NSDI and incorporates tunneling, rescattering and  energy conservation in this process, and the consequences of three-dimensional phase space. In this sense, its results provide a benchmark. An example that intricate structures may be created by these simple ingredients is provided by the momentum correlations presented in Figs.~\ref{fig1}(c,d) and \ref{fig:clct}(c,d): the ring-shaped  populations may suggest the action of a repulsive force, which actually is not there.

It seems that the most important ingredient that is missing from the present analysis is the interaction of the electrons in the intermediate state and the final states with the ion. To some very elementary extent, this is accounted for if we employ the three-body contact interaction for the crucial electron-electron interaction $V_{12}$, while it is definitely not when we take the electron-electron Coulomb interaction \cite{JMO03}. In reality, the presence of the ion will shield the fundamental electron-electron Coulomb repulsion to some extent, which is taken into account in an extreme fashion by the contact interaction. This argument is supported by the good agreement of classical-trajectory (CT) \cite{beijing2} calculations with both the experimental data and the results of our most rudimentary model, since these calculations include all particle interactions at any stage of the process. The particular importance of the ion is also surmised in a recent comparison of the experimental transverse electron-ion correlation with an $S$-matrix calculation \cite{bede}.

The excellent agreement between the results of our quantum-mechanical $S$-matrix calculations and those of the corresponding classical model of Section \ref{classicalmodels} can be invoked to justify such a classical calculation from the outset, provided the parameters are sufficiently far above the classical threshold. This has recently been done in a computation of NSDI by a few-cycle laser pulse as a function of the carrier-envelope phase \cite{lf}. The agreement also lends additional credit to the three-dimensional CT results in the regime sufficiently well inside the classical realm. A corresponding agreement between quantum and classical results has also been observed in the context of one-dimensional model calculations \cite{eberly}. We can make contact with such models by restricting the transverse momenta to values near zero. Of course, recent measurements of NSDI at and below the classical threshold \cite{eremina} are outside the reach of the classical approach.

\begin{acknowledgments} We thank S.V. Popruzhenko and H. Rottke for useful discussions and
D.B. Milo\v{s}evi\'c for providing subroutines. In particular, we 
are highly indebted to
E. Lenz for his help with the code. This work was supported in part by the 
Deutsche Forschungsgemeinschaft.
\end{acknowledgments}


\begin{thebibliography}{99}
\bibitem{pp} The ponderomotive potential of a laser field with vector potential $\vecA(t)$ is defined as the time-averaged wiggling energy $e^2\langle \vecA(t)^2\rangle_t/(2m)$ of an electron in its presence.
\bibitem{hoeberly} When we speak of a ``classically allowed region'' or a ``classical threshold'', this is supposed to be understood as defined, in relation to Eqs.~(\ref{modseq}) at the end of Sec.~\ref{secallow}. Whether description of NSDI genuinely requires quantum mechanics, has recently been questioned \cite{pjho}.
\bibitem{newfoot}Near some particular laser intensities, i.e., near the so-called channel closings, constructive interference may result in a substantial contribution of the longer orbits; see Ref.~\cite{popgorchaclo}.
\bibitem{footnote} It is easy to see from the saddle-point equations (\ref{seq}) that they are incompatible with $\vecp_1=\veck$.
\bibitem{footn} In quantum mechanics, this case is more subtle. See Ref.~\cite{nsdiuni} for a discussion of the corresponding quantum orbits.
\bibitem{finalfoot}In the one-dimensional classical calculations of Ref.~\cite{pjho}, back-to-back emission has been found to be associated with ``speed-up'' collisions, where the returning electron is accelerated by the laser field into the recollision.
\end{thebibliography}
\end{document}